\documentclass[prb,twocolumn,aps,amsmath,amssymb]{revtex4-2}

\UseRawInputEncoding

\usepackage{xcolor}

\usepackage{graphicx}

\usepackage{times}
\usepackage{textcomp}
\usepackage{epstopdf}
\usepackage{float}
\usepackage{bm,mathrsfs}
\usepackage[dvips]{epsfig}
\usepackage{amsmath}
\usepackage{amssymb}
\usepackage{amsfonts}

\usepackage{enumerate}
\usepackage{hhline}
\usepackage{threeparttable}
\usepackage{multirow}
\usepackage{supertabular}
\usepackage{pslatex}
\usepackage{tabularx}
\usepackage{graphicx}
\usepackage{dcolumn}
\usepackage{bm}

\usepackage[colorlinks]{hyperref}

\begin{document}

\title{Longitudinal (curvature) couplings of an $N$-level qudit to a superconducting resonator
at the adiabatic limit and beyond}

\author{Rusko Ruskov, Charles Tahan}

\affiliation{Laboratory for Physical Sciences, 8050 Greenmead Dr., College Park, MD 20740}

\email{charlie@tahan.com, ruskovr@lps.umd.edu}

\date{\today}

\begin{abstract}

Understanding how and to what magnitude solid-state qubits couple to
metallic wires is crucial to the design of quantum systems such as quantum computers.
Here, we investigate the coupling between a multi-level system, or qudit, and a
superconducting (SC) resonator's electromagnetic field,
focusing on the interaction involving both the transition and diagonal dipole
moments of the qudit. Specifically, we explore the effective dynamical (time-dependent)
longitudinal coupling that arises when a solid-state
qudit
is adiabatically modulated at
small gate frequencies and amplitudes, in addition to a static dispersive interaction with the SC resonator.
For the first time, we derive Hamiltonians describing the longitudinal multi-level interactions
in a general dispersive regime, encompassing both dynamical longitudinal and dispersive interactions.
These Hamiltonians smoothly transition between their adiabatic values,
where the couplings of the n-th level are proportional to the
level's energy curvature
concerning a qudit gate voltage, and the substantially larger dispersive values,
which occur due to a resonant form factor.

We provide several examples illustrating the transition from adiabatic to dispersive coupling
in different qubit systems, including the charge (1e DQD) qubit, the transmon,
the double quantum dot singlet-triplet qubit, and the triple quantum dot exchange-only qubit.
In some of these qubits, higher energy levels play a critical role,
particularly when their qubit's dipole moment is minimal or zero.
For an experimentally relevant scenario involving
a spin-charge qubit with magnetic field gradient
coupled capacitively to a SC resonator,
we showcase the potential of these interactions.
They enable close-to-quantum-limited quantum non-demolition (QND) measurements and
remote geometric phase gates, demonstrating their practical utility in quantum information processing.

\end{abstract}

\maketitle

\section{Introduction}

Effective coupling of a multi-quantum-dot encoded spin qubit to a superconducting (SC) resonator
would be a major step towards establishing  high fidelity qubit quantum measurement as well as
long-range spin-spin interactions on an electronic chip.
Following the success with superconducting qubits
\cite{Wallraff2005PRL,ClelandMartinis2013PRL,Bergeal2010N,Majer2007N},
the usual approach is to establish a transverse (dipole) coupling, $g_{\perp}$, of the qubit to the
quantized e.m. field of the resonator. 
In a dispersive regime (at a qubit-resonator frequency detunning $\Delta$)
an excitation exchange between qubit and resonator is suppressed by a small probability,
$\left( \frac{g_{\perp}}{\Delta}\right)^2$.
Thus, avoiding a direct excitation of the qubit by the resonator in this limit,
a quantum non-demolition (QND) measurement
is possible, with non-QND effects of the order of $\left( \frac{g_{\perp}}{\Delta}\right)^2$.

Despite the
smaller dipole strength and in the presence of
stronger charge noise,
such a dispersive coupling was also successfully introduced
in the QD spin-qubit architectures
\cite{Petersson2012N,Petta2017Science,MiPetta2018N,
Landig-Wallraff-Ensslin-Ihn2018N,Samkharadze-Vandersypen2018Sci,
BottcherYacobyNatComm2022,CorriganHarptRuskovEriksson2023PRAppl}.
Besides the non-QND effects, this approach may suffer from two main issues
(both for SC qubits and QD spin qubits):
(1) qubit
relaxation Purcell enhancement via the resonator due to relative closeness
of the qubit and resonator frequencies in the dispersive regime;
(2) Typically, a large transverse (dipole) coupling
would couple to nearby two-level fluctuators, leading to charge noise.
Also, for QD spin-qubits  a large dipole coupling
arises at regions of a ``charge degeneracy points'' (c.d.p.),
where the gate charge noise is enhanced as well.

Motivated to avoid these issues, we have proposed
\cite{RuskovTahan2017preprint1,RuskovTahan2019PRB99}
qubit-resonator energy curvature couplings in the adiabatic regime for encoded multi-dot spin qubits
(in which the qubit dipole moment is zero or suppressed),
such as the triple quantum dot (TQD) always on exchange only (AEON) qubit and the
double quantum dot singlet-triplet (DQD S-T) qubit at their symmetric operating point (SOP).
(Similar proposals already existed for superconducting qubits
\cite{Kerman2013,Billangeon2015PRB,Didier2015PRL,RicherDiVincenzo2016PRB}).
The non-zero energy curvature (with respect to a gate voltage),
arising from higher levels,
is essentially
the qubit quantum capacitance, $C_q \propto \frac{\partial^2 E_q}{\partial V_m^2}$,
and it introduces two effective QND interactions to the resonator:
\begin{eqnarray}
&& {\cal H}_{\delta\omega}/\hbar = \delta\omega\, \sigma_z \hat{a}^\dagger \hat{a},
\label{qubit_curvature_Hamiltonian-disp-like}
\\
&& {\cal H}_{\parallel}/\hbar =
\left[\tilde{g}_{\parallel}\, \sigma_z + \tilde{g}_{\rm av}\right](\hat{a}^\dagger + \hat{a})\, \cos(\omega_m t)
\label{qubit_curvature_Hamiltonian-lonitudinal}
\end{eqnarray}
The first one is an always on dispersive coupling
with strength $\delta\omega \propto g_0^2\, C_q$,
and the second one is a dynamical longitudinal coupling
with strength $\tilde{g}_{\parallel} \propto g_0\, C_q\, \tilde{V}_m$,
that arises when a suitable qubit gate voltage
is time modulated, $\sim \tilde{V}_m\, \cos(\omega_m t)$,
at or around the resonator frequency,
in an adiabatic regime of small frequencies
and modulation amplitudes.
Both couplings are proportional to the qubit energy curvature $C_q$,
while $\tilde{g}_{\parallel}$ can be switched on/off
by the external gate voltage modulation.
In the above we have introduced qubit-resonator ``bare'' dipole coupling,
$g_0 \equiv \alpha_c\,\frac{\omega_r}{2}\, \sqrt{\frac{Z_r}{\hbar/e^2}} \ll \omega_r$
(with $\alpha_c$ and $Z_r$ being the resonator-to-dot lever arm and resonator impedance,
respectively),
see below and
Ref.~\onlinecite{Childress2004}.
These couplings were derived in a ``soft-field'' approach \cite{RuskovTahan2017preprint1,RuskovTahan2019PRB99}
at the SOP,
essentially in the adiabatic regime
\begin{equation}
\omega_r \ll E_{\rm gap} \sim U_{\rm charge} ,
\label{adiabaticity_gap}
\end{equation}
since
the typical energy gap to compare
for these encoded qubits
is the dot's charging energy, $U_{\rm charge} \geq 200\, {\rm GHz}$,
(in the absence of a qubit transverse dipole moment at the SOP,
see below and Ref.~\onlinecite{RuskovTahan2019PRB99}).
While at the SOP
it was proposed to use these couplings to perform
a QND quantum measurement \cite{RuskovTahan2019PRB99} as well as
N-qubit geometric phase gates \cite{HarveyYacoby2018PRB,RuskovTahan2021PRB103},
the high-performance regime of these operations (reaching infidelities of $10^{-3}$)
was shown to be limited
either by
small quantum capacitances
at the SOP (for realistic experimental parameters)
or
by a necessity to work at a small ratio
\cite{RuskovTahan2021PRB103},
$\sim \frac{\delta\omega}{\tilde{g}_{\parallel}}$.

Despite  the SOP regime, it is worth to study also
the regime of dots' detunnings closer to the c.d.p.
Indeed, the ratio of the curvatures at c.d.p vs. SOP
for encoded spin qubits
can reach high values\cite{RuskovTahan2019PRB99,RuskovTahan2021PRB103}:
\begin{equation}
\frac{C_q^{\rm c.d.p.}}{C_q^{\rm SOP}} \sim \frac{U_{\rm charge}^3}{16 t_c^3} \sim 10^2 - 10^4
\label{ratio_curvatures}
\end{equation}
for typical charging energies,
$U_{\rm charge} \sim 200 - 300\, {\rm GHz}$, and tunnelings, $t_c \sim 5 - 10\, {\rm GHz}$
(in frequency units).
%
At the c.d.p. a typical qubit level splitting (with an attached to it transition dipole moment)
is of the order of $\omega_q \sim 2 t_c \ll U_{\rm charge}$,
and the adiabatic regime,
$\omega_r \ll E_{\rm gap} \sim 2 t_c$,
requires much smaller resonator frequencies.
Recently, both the dispersive and dynamical longitudinal couplings
have been observed in the adiabatic regime for a charge qubit coupled
to SC resonator \cite{CorriganHarptRuskovEriksson2023PRAppl}.
%
As the qubit and resonator are highly detuned ($\Delta \approx \omega_q$) the Purcell effect
will be strongly suppressed.
In this adiabatic regime
the relevant 
dispersive  coupling
$\delta\omega \propto \frac{g_{\perp}^2}{\omega_q}$ (see below),
will be   also
suppressed with respect to the
usual dispersive coupling (since $\omega_q \gg \Delta\approx 10 g_{\perp}$).
However, this suppression may be compensated via much stronger dynamical longitudinal coupling.
Estimations give for their ratio
(this ratio is independent of the qubit frequency detuning, see below):
\begin{equation}
\frac{\tilde{g}_{\parallel}}{\delta\omega} =  \frac{e \tilde{V}_m}{2 \hbar g_0}
\equiv \frac{\tilde{V}_m}{\alpha_c V_{\rm vac}}
\sim 12 - 120 ,
\label{ratio_classical-quantum}
\end{equation}
where $V_{\rm vac} = \frac{\hbar \omega_r}{e}\, \sqrt{\frac{Z_r}{\hbar/e^2}}$
is the amplitude of
zero-point (vacuum) fluctuations of the resonator.
Then, the overall quantum
measurement rate (with modulation) \cite{RuskovTahan2019PRB99}
in the adiabatic regime
can be of the order or stronger
than the usual measurement rate of the dispersive regime.

Since the relevant energy gap around the c.d.p., $E_{\rm gap} \sim 2 t_c$, is relatively small, the adiabatic regime
may not be fully satisfied (e.g., when the qubit detuning is in the usual dispersive regime).
It is then desirable to derive effective Hamiltonians of the type
of Eqs.~(\ref{qubit_curvature_Hamiltonian-disp-like})
and
(\ref{qubit_curvature_Hamiltonian-lonitudinal}),
which would be relevant both in the adiabatic regime
and in the much less detuned dispersive regime.

In this paper we perform the task for a general n-level system (qudit),
such that the above effective Hamiltonians obtain contributions from many levels,
not just from the qubit levels, in a general dispersive regime.
First, in Sec.~\ref{SecII-dipole} we introduce the usual
dipole interaction of an n-level atom to a quantized or classical
electromagnetic field.
In Sec.~\ref{SecIII-adiabatic}
we derive the corresponding effective Hamiltonians in the adiabatic regime,
using the ``soft field'' approach.
Then, in Sec.~\ref{SecIV-A-perturbative}
similar effective Hamiltonians are derived
in an approach based on a time-dependent
quantum mechanical (QM) perturbation theory, extending them to a regime when
resonator frequency dependence is not neglected.
In Sec.~\ref{SecIV-B0-frequences} we obtained
more general result, when the modulation frequency
is different from the resonator frequency.
In Sec.~\ref{SecIV-B-Schrieffer-Wolff} and in Appendix~\ref{app-A: Schrieffer-Wolff-PT}
the equal frequency case is re-derived in a  formal
time-dependent Schrieffer-Wolff transformation approach.
In Sec.~\ref{SecIV-C-polarizability} we further extend these
results by adding an atom's polarizability contribution
to the effective Hamiltonians, ${\cal H}_{\delta\omega}$ and
${\cal H}_{\parallel}$, which is of the same order ($\sim g_0^2$)
as the above second order PT contributions.
This
provides
the most general expressions for
the effective Hamiltonians, ${\cal H}_{\delta\omega}$ and
${\cal H}_{\parallel}$,
of Eqs.~(\ref{qubit_curvature_Hamiltonian-disp-like}) and
(\ref{qubit_curvature_Hamiltonian-lonitudinal}).
The polarizability contribution for each level
is recast to the atom's energy level curvatures plus
a sum over atom's transition dipole matrix elements
via a ``low-energy QM sum rule'' (Appendix \ref{app-B: Low-energy-sum-rule}).
This allows us to perform the adiabatic limit in the general
effective Hamiltonians,
Sec.~\ref{SecV-limiting-regimes} A,
confirming the ``soft-field'' results.
We also consider the case of zero polarizability
in Sec.~\ref{SecV-limiting-regimes} B, relevant to QD spin-qubits,
as well as the dispersive regime in Sec.~\ref{SecV-limiting-regimes} C.
Out of the adiabatic regime, the second approach will provide an $\omega_r$ dependence
of the effective Hamiltonian strengths,
$\delta\omega$,
$\tilde{g}_{\parallel}$,
which approaches that of the usual dispersive regime
in the appropriate limit.

In Sec.~\ref{SecVI-examples} A, B, C, D,
we apply the general theory to several cases of interest, including
a DQD charge qubit,
a Transmon, a DQD S-T qubit,
and TQD exchange only qubit
(in Appendix \ref{app-C: Dipole-m-e-TQD-qubit} we calculated the
dipole matrix elements for the TQD system).
Finally, in Sec.~\ref{SecVI-examples} E,
we consider implications of the theory for (continuous) quantum measurements
on the example of a charge qubit \cite{CorriganHarptRuskovEriksson2023PRAppl}
and a spin-charge qubit with magnetic field gradient \cite{HuLiuNori2012PRB,MiPetta2018N}.
Implications for Geometric quantum gates \cite{HarveyYacoby2018PRB,RuskovTahan2021PRB103}
are briefly mentioned in Sec.~\ref{SecVI-examples} F.

\section{Dipole interactions of an $n$-level system with e.m. fields}
\label{SecII-dipole}
\begin{figure} [t!] 
    \centering
     \includegraphics[width=0.46\textwidth]{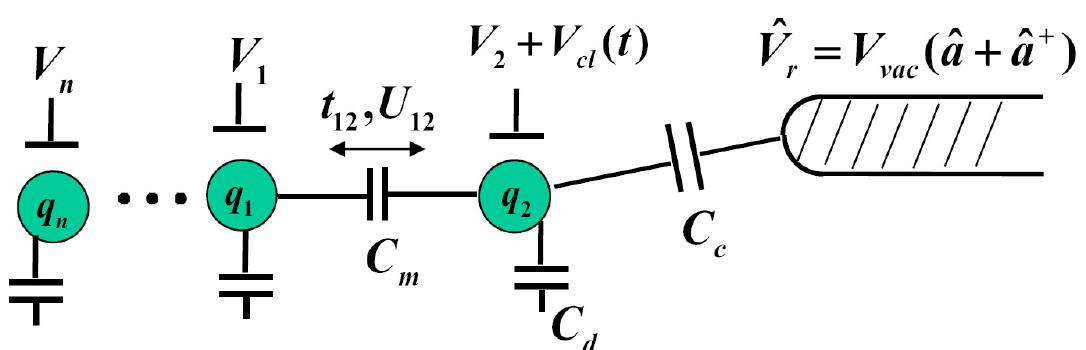}
        \caption{
        A multi-quantum dot system is coupled capacitively to a SC resonator via dot 2.
        The voltage variations at the dot 2 can be quantum, $\delta V_2 = \alpha_c \hat{V}_r$,
        or classical, $\delta V_2 = V_{\rm cl}(t) = \tilde{V}_m \cos (\omega_m t)$;
        $\alpha_c \simeq \frac{C_c}{C_c + C_d}$ is the resonator-to-dot lever arm,
        where $C_c$ and $C_d$ are dot's coupling capacitances.
        }
        \label{fig:1}
\end{figure}
One starts with a qudit plus resonator Hamiltonian including the dipole interaction
with the e.m. field of the resonator,
$\hat{H}_{\rm dipole}(t)\equiv -\hat{\overrightarrow{d}} \cdot \delta\overrightarrow{E}_\lambda(t)$
(see Fig.~\ref{fig:1} for a multi-quantum dot system):
\begin{eqnarray}
&&
{\cal H}_{\rm tot}  =  \sum_k  E_k \, |k\rangle \langle k |  + \hbar \omega_r^{(\lambda)} \hat{a}^\dagger \hat{a}
+ \hat{H}_{\rm dipole}(t)
\\
&&
\hat{H}_{\rm dipole}(t) =
- \sum_{k,j} \overrightarrow{d}_{kj} \cdot \delta \overrightarrow{E}_\lambda(t)
\label{system_em_interaction}
\end{eqnarray}
Here $E_k$ is the energy of the $k$-th level, $\omega_r^{(\lambda)}$ is the frequency
of a single resonator mode $(\lambda)$,
and the e.m. field applied to the system has quantum and classical components:
\begin{equation}
\delta\overrightarrow{E}_\lambda(t) = \overrightarrow{E}_{\lambda,\rm vac} (\hat{a} + \hat{a}^\dagger)
+ 2 \overrightarrow{E}_{\rm cl}\, \cos(\omega_m t) .
\label{em-field-quant-classical}
\end{equation}
The quantum field of the resonator has a normalized amplitude, $\overrightarrow{E}_{\lambda,\rm vac}$
(chosen real, $\overrightarrow{E}_{\lambda,\rm vac} = \overrightarrow{E}_{\lambda,\rm vac}^*$,
see e.g. \cite{MilburnWalls-book2008}).
The classical field amplitude, $\overrightarrow{E}_{\rm cl}$, may come from an applied gate voltage
to the dot system (see Fig.~\ref{fig:1}) which is generally time-dependent.
The dipole matrix elements $\overrightarrow{d}_{kj}$  are either off-diagonal (transition dipole m.e.)
or diagonal ($j=k$).
The latter are generally non-zero for a multi-dot system where space parity is not a good quantum number.
In what follows we skip the index of the resonator mode so that $\omega_r^{(\lambda)} \equiv \omega_r$
(the results can be easily extended to a multi-mode case).

It is often more convenient to deal with applied voltages to the multi-dot system instead of the corresponding
electric fields.
As an example, consider a DQD with a gate voltage at the right dot, $V_2$.
Then $E_x = \frac{V_2}{l_x}$,
where $l_x$ is the distance between dots 1 and 2, Fig.~\ref{fig:1}.
If $\delta V_2$ is induced by the quantized voltage of the resonator \cite{Childress2004},
$\hat{V}_r = V_{\rm vac}\, (a + a^\dagger)$
then the corresponding field amplitude is:
\begin{equation}
E_{x, vac} = \alpha_c \frac{V_{\rm vac}}{l_x} \equiv \frac{2 \hbar}{l_x e}\, g_0
\label{vac_field}
\end{equation}
where $\alpha_c \simeq \frac{C_c}{C_c + C_d}$  is the lever arm of dot 2 to the resonator,
and $g_0 \equiv \alpha_c \frac{e V_{\rm vac}}{2\hbar}$
is the DQD ``bare'' coupling to the resonator,
introduced above.

The dipole coupling in Eq.~(\ref{system_em_interaction}) is then
$g_{kj} \equiv d_{x,kj} E_{x, vac} = \frac{(\hat{x})_{kj}}{l_x}\, 2 g_0$.
By introducing also a classical voltage variation to dot 2
[$\delta V_2 = V_{\rm cl}(t) \equiv \tilde{V}_m\,\cos(\omega_m t)$],
Eqs.~(\ref{system_em_interaction}) and (\ref{em-field-quant-classical})  can be re-written as:
\begin{equation}
\hat{H}_{\rm dipole}(t) = \sum_{j,k}\, \hbar g_{kj}\, |k \rangle  \langle j|
\left\{ \left[\hat{a}+ b_{\rm cl}(t) \right] + \left[ \hat{a}^\dagger + b^{*}_{\rm cl}(t) \right] \right\},
\label{V-dipole-transverse-coupl}
\end{equation}
where $b_{\rm cl}(t) \equiv \overline{b}_{\rm cl}\, e^{-i \omega_m t}$
is a classical field
with amplitude
\begin{equation}
\overline{b}_{\rm cl} = \frac{e \tilde{V}_m}{4 \hbar g_0} = \frac{\tilde{V}_m}{\alpha_c V_{\rm vac}}.
\label{b_classical}
\end{equation}
From Eqs.~(\ref{V-dipole-transverse-coupl}) and (\ref{b_classical}) it is clear
that the small parameters of the problem are the energies associated with  the quantized and classical
voltage variations, $e V_{\rm vac}$ and $e \tilde{V}_m$ (relative to the qubit energy scale, $E_q$).

\section{$n$-level effective Hamiltonians in the ``soft-field'' (adiabatic) limit}
\label{SecIII-adiabatic}

\begin{figure} [t!] 
    \centering
     \includegraphics[width=0.4\textwidth]{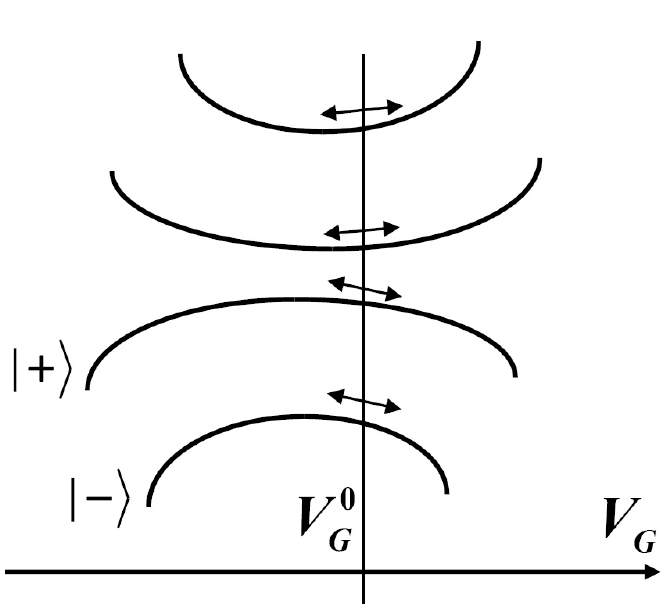}
        \caption{
        A generic multi-level system where the energy levels depend on a voltage parameter $V_G$.
        By (Taylor) expanding the energy levels to second order around a working point $V_G^0$, where
        the voltage variation $\delta V_G = V_{\rm vac} (\hat{a} + \hat{a}^{\dagger}) + V_{\rm cl}(t)$,
        obtains quantum and classical part, allows to derive effective adiabatic interactions
        of the $n$-level system to a super-conducting resonator, Eqs.(\ref{qudit-dispersive-like}) and
        (\ref{qudit-dynamical-longitudinal}),
        Sec.~\ref{SecIII-adiabatic},
        that are proportional to the levels' energy curvatures
        (quantum capacitances),
        shown with $\leftrightarrow$ for each level.
        These effective interactions imply adiabaticity condition,
        $\omega_r \ll E_{\rm gap}$, where a typical energy gap, $E_{\rm gap}$ is either the qubit splitting
        or some higher energy transition, see Eq.~(\ref{adiabaticity_gap}).
        The adiabatic interactions, Eqs.(\ref{qudit-dispersive-like}) and
        (\ref{qudit-dynamical-longitudinal}) are
        re-derived in Secs.~\ref{SecIV} and \ref{SecV-limiting-regimes}
        in a perturbation theory that also allows to go beyond the adiabatic regime.
        }
        \label{fig:2}
\end{figure}
Consider an adiabatic limit where $\omega_r \ll |\omega_{jk}| \equiv |\omega_j - \omega_k|$ for any energy levels,
and also that the voltage variations at the dot 2 (quantum and classical),
$\alpha_c \hat{V}_r$, $V_{\rm cl}(t) \equiv \tilde{V}_m\, \cos(\omega_m t)$
are considered small
(or ``soft-field'')
so that photon excitations of the qudit are highly suppressed, Fig.~\ref{fig:2}.
(Also, in this limit it is assumed that phonon assisted excitations
are
suppressed, and therefore neglected,
compare with Refs.~\cite{GullansPettaTaylor2016PRL,StehlikGullansPetta2016PRX}).
Assuming
that the qudit energy levels depend on the dot gate voltage,
one can Taylor expand the energy levels
to second order in the gate voltage variation,
$\delta V_G = V_{\rm vac}\, (\hat{a} + \hat{a}^\dagger) + V_{\rm cl}(t)$,
that contains a quantum and classical part
\footnote{In an early work \cite{Ruskov-Higgs1987} a Taylor expansion in light Higgs fields
allowed to obtain effective interactions to light hadrons without calculating
Feynman diagrams. }:
\begin{eqnarray}
&& E_k(V_G) = E_k(V_G^0) + \frac{\partial E_k}{\partial V_G}\, \delta V_G
+ \frac{1}{2}\, \frac{\partial^2 E_k}{\partial V_G^2}\, \delta V_G^2 ,
\nonumber\\
&&  {} = E_k(V_G^0) + Q^{(k)}_d \, \delta V_G + \frac{1}{2}\, C^{(k)}_d\, \delta V_G^2 .
\label{Taylor-expand}
\end{eqnarray}
In the second row of Eq.~(\ref{Taylor-expand})
$Q^{(k)}_d \equiv \frac{\partial E_k}{\partial V_G}$ is a (quasi)charge
%
\footnote{For a DQD charge qubit, $Q^{(k)}$ can be recast to the excess charge
on the right dot, see
Ref.~\onlinecite{Dicarlo2004PRL}.
}.
It generates a static longitudinal interaction
\cite{Billangeon2015PRB,RicherDiVincenzo2016PRB,RuskovTahan2017preprint1,RuskovTahan2019PRB99},
see also Eq.~(\ref{static_longitudinal_transverse}) below.
The coefficient in the second order term,
$C^{(k)}_d \equiv\frac{\partial^2 E_k}{\partial V_G^2}$,
is the quantum capacitance of the $k$-th level.
%
Substituting for $\delta V_G$ in the second order
one obtains the following
effective qudit-resonator
adiabatic  Hamiltonians:
\begin{equation}
{\cal H}_{\delta\omega}/\hbar
= 4 \hbar g_0^2 \frac{\partial^2 \hat{\Sigma}}{\partial V_G^2}\, (\hat{a}^\dagger\hat{a} + \frac{1}{2})
\label{qudit-dispersive-like}
\end{equation}
and
\begin{equation}
{\cal H}_{\parallel}/\hbar
= 2 g_0 (e \tilde{V}_m) \frac{\partial^2 \hat{\Sigma}}{\partial V_G^2}\, (\hat{a} + \hat{a}^\dagger)\, \cos(\omega_m t) ,
\label{qudit-dynamical-longitudinal}
\end{equation}
which replace the original dipole interactions, Eq.~(\ref{V-dipole-transverse-coupl}),
in the adiabatic regime.
Here, $\hat{\Sigma} \equiv {\rm diag}\{E_1,E_2,\dots,E_k,\dots\}$ is the qudit eigenenergy matrix.
For a qubit the above expressions coincide with that of
Refs.~\cite{RuskovTahan2017preprint1,RuskovTahan2019PRB99}.

The effective interactions,
Eqs.~(\ref{qudit-dispersive-like}) and (\ref{qudit-dynamical-longitudinal}) can be used to perform
(continuous) quantum measurements on a qudit via measuring the photo-current of a nearby
superconducting (SC) resonator \cite{RuskovTahan2017preprint1,RuskovTahan2019PRB99}
(see also recent paper, Ref.~\cite{SteinmetzSiddiqiJordan2022PRA105}).
The simplest example could be a DQD singlet-triplet qubit, which has three relevant levels
and the qubit subspace has no dipole moment for any DQD detunning
(see Sec.~\ref{SecVI-examples} C below).
Another interesting example relevant to the experiment is
a spin-charge qubit where a DQD charge qubit (1e) with a micromagnet-induced gradient
magnetic field between the two dots creates effective dipole coupling to
the resonator
\cite{MiPetta2018N,HuLiuNori2012PRB}
(see Sec.~\ref{SecVI-examples} E below).
Since, in the relevant experiments the adiabatic conditions
may not be
fulfilled, we develop an alternative derivation
of the
effective interactions of the type of
(\ref{qudit-dispersive-like}) and (\ref{qudit-dynamical-longitudinal}),
extending their range of applicability.

\section{$n$-level effective Hamiltonians from a perturbation theory}
\label{SecIV}

\subsection{Derivation via second order time-dependent perturbation theory}
\label{SecIV-A-perturbative}
The dipole interaction Eq.~(\ref{V-dipole-transverse-coupl}) is a time-dependent perturbation.
One (heuristic) way to deal with it is to apply a time-dependent perturbation theory (PT).
By treating the quantized field in the Heisenberg picture
one arrives at time-dependent field operators, $\hat{a}(t) =  \hat{a}\, e^{-i \omega_r t}$,
that will be treated semi-classically, i.e., on the same ground as
the classical fields \cite{Haken-book1976,Ruskov-MarchMeeting2019-talk},
$b_{\rm cl}(t) \equiv \overline{b}_{\rm cl}\, e^{-i \omega_m t}$,
in Eq.~(\ref{V-dipole-transverse-coupl}).
%
The
Hamiltonian is then $\hat{H}(t) = \hat{H}_0 + \hat{V}(t)$
with
a time-dependent perturbation atom's matrix element of the form:
\begin{equation}
V_{kl}(t) = \hbar g_{kl}\, (\hat{a} \, e^{-i \omega_r t}
+ \overline{b}_{\rm cl} \, e^{-i \omega_m t} ) + {\rm h.c.}
\label{1st-order-dipole-m-e}
\end{equation}
%

Working in the interaction picture the Hamiltonian becomes
$\bar{V}(t) = e^{i \frac{\hat{H}_0}{\hbar}t}\, \hat{V}(t) \, e^{-i \frac{\hat{H}_0}{\hbar}t}$.
To obtain an effective Hamiltonian in a non-resonant (dispersive) regime,
$|\omega_k - \omega_l - \omega_r| \gg g_{kl}$ for any $k,l$,
and
to effectively eliminate the fast oscillating components of $\bar{V}(t)$,
one performs a unitary transform to the system's state,
$\tilde{\rho} = e^{-i Q(t)}\, \rho \, e^{i Q(t)}$,
and a corresponding Hamiltonian transform
\begin{equation}
\tilde{V}(t) = e^{-i Q(t)}\, \bar{V}(t) \, e^{i Q(t)}
- i\hbar\, e^{-i Q(t)}\, \frac{\partial}{\partial t} e^{i Q(t)} ,
\label{Hamiltonian_transform}
\end{equation}
requiring that the fast oscillating components of $\bar{V}(t)$ to be removed.
Assuming that $\bar{V}(t) \propto \lambda$ has a small parameter (e.g., the field amplitudes),
one can expand $Q(t)$ and $\tilde{V}(t)$ in a perturbation series:
\begin{eqnarray}
&& Q(t) = Q_1(t) + Q_2(t) + \cdots  ,
\\
&& \tilde{V}(t) = \tilde{V}_0(t) + \tilde{V}_1(t) + \tilde{V}_2(t) + \cdots
\label{perturbation series}
\end{eqnarray}
where $Q_n(t),\tilde{V}_n(t) \propto \lambda^n$.
Using the Baker-Hausdorff expansion:
$e^{-i Q}\, \bar{V} \, e^{i Q} = \bar{V}(t) - i\, \left[Q,\bar{V}(t)\right]
- \frac{1}{2}\, \left[Q,\left[Q,\bar{V}\right]\right] + \cdots$,
one substitutes in it
the
perturbative series to obtain the following equations
collected at a given power of $\lambda$:
\begin{eqnarray}
&& \tilde{V}_0(t) = 0
\label{lambda_0}
\\
&& \tilde{V}_1(t) = \bar{V}(t) + \hbar \frac{\partial Q_1(t)}{\partial t}
\label{lambda_1}
\\
&&  \tilde{V}_2(t) = -\frac{i}{2} \left[Q_1,\bar{V}(t)\right]
                 -\frac{i}{2} \left[Q_1,\tilde{V}_1(t)\right] + \hbar \frac{\partial Q_2(t)}{\partial t}
\label{lambda_2}
\\
&&  \ldots
\nonumber
\end{eqnarray}
One requires that $\tilde{V}_1(t) = 0$ in the non-resonant dispersive case
(in the original frame $\bar{V}(t)$ is highly oscillating, so one effectively remove it here).
Then, the solution of Eq.~(\ref{lambda_1}) takes the standard form
of a time dependent PT \cite{LandauLifshitz-book2003}:
\begin{equation}
Q_{1,kl}(t) = -\frac{1}{\hbar} \, \int_{-\infty}^t dt' \, e^{i \omega_{kl} t'} \, V_{kl}(t')
\label{Q1-term} ,
\end{equation}
where $\omega_{kl} \equiv (E_k - E_l)/\hbar$.
The effective Hamiltonian can be obtained from the second order
diagonal term, $\tilde{V}_{2,kk}(t)$, thus requiring that the
off-diagonal elements are zero, eliminating highly-oscillating terms.
[Analogously to the above, one requires that $\tilde{V}_{2,kl}(t) = 0$ for $k\neq l$, while $Q_{2,kk}=0$.
Thus, $Q_{2,kl}$ has only off-diagonal elements and is obtained from the equation:
$\hbar \frac{\partial Q_{2,kl}(t)}{\partial t} = \frac{i}{2} \left[Q_1, \bar{V}(t)\right]_{kl}, \ k\neq l$,
which can be used for higher order calculations.]

In the lowest order, from Eq.~(\ref{lambda_2}) with $\tilde{V}_1(t) = 0$, one then obtains:
\begin{equation}
\tilde{V}_{2,kk}(t) \equiv U_{kk}^{\rm eff}(t) =
- \frac{i}{2} \, \sum_l \, \left[ Q_{1,kl}(t) \bar{V}_{lk}(t) - \bar{V}_{kl}(t) Q_{1,lk}(t) \right]
\label{commutator}
\end{equation}
Calculating the commutator for equal frequencies, $\omega_m = \omega_r$
(In this paper we mainly focus on this resonant case,
with some exceptions, see below),
and neglecting the fast-oscillating (contra-rotating) terms
\footnote{In this approach the counter-rotating (fast-oscillating) terms
are that of the first order $\sim g_0$, Eq.~(\ref{1st-order-dipole-m-e}),
as well as the off-diagonal in atomic index terms
of the second order $\sim g_0^2$, in Eq.~(\ref{lambda_2})
}
%
%
in a rotating wave approximation (RWA),
one obtains an effective adiabatic/dispersive Hamiltonian (diagonal in the atomic index),
\begin{eqnarray}
&& {\cal H}_{\rm eff}/\hbar = \sum_k U_{kk}^{\rm eff}/\hbar \,\, |k\rangle \langle k |
\nonumber\\
&& \quad\ \ \, { } = \sum_{k,l} \ \left\{ \left[ r_{kl}(\omega_r) - r_{lk}(\omega_r) \right] \right.
\nonumber\\
&& \qquad\quad \left.
\times \left( \hat{a}^\dagger \hat{a}
+ \hat{a}\,\, b^*_{\rm cl}(t)
+ \hat{a}^\dagger \, b_{\rm cl}(t) \right)
\right.
\nonumber\\
&& \quad\ \ \ \left. { } + \left[ r_{kl}(\omega_r) - r_{lk}(\omega_r) \right]  \, |\overline{b}_{\rm cl}|^2
+ r_{kl}(\omega_r)  \right\} \, |k\rangle \langle k | ,
\label{H-eff}
\end{eqnarray}
where
\begin{equation}
r_{kl}(\omega_r) \equiv  \frac{|g_{kl}|^2}{\omega_{kl} - \omega_r} .
\label{chi}
\end{equation}

The effective Hamiltonian Eq.~(\ref{H-eff}) contains a ``dispersive-like'' coupling,
$D_k(\omega_r) \, \hat{a}^\dagger \hat{a} \, |k\rangle \langle k |$,
and  a dynamical longitudinal coupling,
$L_k(\omega_r) \, (\hat{a} + \hat{a}^\dagger) \, |k\rangle \langle k | \, \cos(\omega_r t)$,
of the $k$-th level to the resonator,
where
\begin{eqnarray}
&& D_k(\omega_r) \equiv \sum_{l\neq k} \, \left[ r_{kl}(\omega_r) - r_{lk}(\omega_r) \right]
\label{D_k}
\\
&&  L_k(\omega_r) \equiv
\sum_{l\neq k} \, 2 \overline{b}_{\rm cl}\, \left[ r_{kl}(\omega_r) - r_{lk}(\omega_r) \right]
\label{L_k}
\end{eqnarray}
are given by a sum of terms, $r_{kl}(\omega_r) \sim g_0^2$, over dipole transitions, $k \to l,\ l\neq k$.
%
Introducing a drive-independent frequency shift, $\delta\omega_k^0(\omega_r)$,
and a drive-dependent one, $\delta\omega_k^{\overline{b}_{\rm cl}}(\omega_r)$,
\begin{eqnarray}
&& \delta\omega_k^0(\omega_r) \equiv \sum_{l\neq k} r_{kl}(\omega_r)
\label{drive-independent-shift}
\\
&& \delta\omega_k^{\overline{b}_{\rm cl}}(\omega_r) \equiv \sum_{l\neq k} {\overline{b}_{\rm cl}}^2 \,
\left[ r_{kl}(\omega_r) - r_{lk}(\omega_r) \right] ,
\label{drive-dependent-shift}
\end{eqnarray}
one gets from Eq.~(\ref{H-eff}):
\begin{eqnarray}
&& {\cal H}_{\rm eff}/\hbar =
\sum_{k} \ \left\{ D_k(\omega_r) \, \hat{a}^\dagger \hat{a}
+ L_k(\omega_r) \, (\hat{a} + \hat{a}^\dagger) \, \cos(\omega_r t)
\right.
\nonumber\\
&& \qquad\qquad
{ } + \left. \delta\omega_k^0(\omega_r)
+ \delta\omega_k^{\overline{b}_{\rm cl}}(\omega_r)
\right\}
 \, |k\rangle \langle k | ,
\label{H-eff-1}
\end{eqnarray}
The second order PT effective Hamiltonian, Eq.~(\ref{H-eff}) [or Eq.~(\ref{H-eff-1})],
is one of the main results of this paper, particularly relevant for QD spin qubits.
In Sec.~\ref{SecIV-C-polarizability} we will show, however, that in general,
${\cal H}_{\rm eff}$ may obtain additional atom's polarizability
contributions of the same order
$\sim g_0^2$.
The latter contributions are $\omega_r$-independent
and are important to obtain the correct adiabatic limit
(Sec.~\ref{SecV-limiting-regimes} A).

\subsection{The case of different frequencies, $\omega_m \neq \omega_r$}
\label{SecIV-B0-frequences}

While in this paper we consider mostly the resonant case
 of equal frequencies,
we provide here the result for different frequencies for the sake of further reference.
The above effective Hamiltonian, Eq.~(\ref{H-eff-1})
is then transformed to:
\begin{eqnarray}
&& \tilde{{\cal H}}_{\rm eff}/\hbar =
\sum_{k} \ \left\{ D_k(\omega_r) \, \hat{a}^\dagger \hat{a}
+ \tilde{L}_k(\omega_r,\omega_m) \, (\hat{a} + \hat{a}^\dagger) \, \cos(\omega_m t)
\right.
\nonumber\\
&& \qquad\qquad
{ } + \left. \delta\omega_k^0(\omega_r)
+ \delta\omega_k^{\overline{b}_{\rm cl}}(\omega_m)
\right\}
 \, |k\rangle \langle k | ,
\label{H-eff-2-omega_m-omega_r}
\end{eqnarray}
where
\begin{equation}
\tilde{L}_k(\omega_r,\omega_m) \equiv
\sum_{l\neq k} \, \overline{b}_{\rm cl}\, \left[ r_{kl}(\omega_r) - r_{lk}(\omega_r)
+ r_{kl}(\omega_m) - r_{lk}(\omega_m) \right]
\label{L_k-modified}
\end{equation}

\subsection{Alternative derivation via Schrieffer-Wolff transformation}
\label{SecIV-B-Schrieffer-Wolff}
Alternatively (see Appendix~\ref{app-A: Schrieffer-Wolff-PT}),
Eq.~(\ref{H-eff}) can be obtained via a time-dependent Schrieffer-Wolff transformation
(see, e.g., Refs.~\cite{MagesanGambetta2020PRA,GoviaKamal2022PRA})
such that
$${\cal H}_{\rm eff} = U^\dagger(t)\, {\cal H}(t) \, U(t)
- i\hbar \frac{\partial U^\dagger(t)}{\partial t} U(t)$$
with $U(t) = e^{S_1(t)}$ and $S_1(t)^\dagger = -S_1(t)$:
%
\begin{eqnarray}
&& S_1(t) = \sum_{lk}\, \left\{ -\frac{g_{kl}}{\omega_{lk} - \omega_r} \, |k\rangle \langle l|\,
\left(\hat{a}^\dagger + b^*_{\rm cl}(t) \right)  \right.
\nonumber\\
&& \quad\qquad \left. { } + \frac{g_{lk}}{\omega_{lk} - \omega_r} \, |l\rangle \langle k |\,
    \Large(\hat{a} + b_{\rm cl}(t) \Large) \right\} .
\label{S_1_solution}
\end{eqnarray}
This particular transformation is a generalization of the time-independent case \cite{Zhu:2013}.

\subsection{Generalization for systems with non-linear voltage dependence}
\label{SecIV-C-polarizability}
In the models of quantum circuits that include QD qubits, the qubit charging energy
is a linear function of the dots' gate voltages in a charge basis
where each charge state corresponds to certain QD's occupation.
(The charge basis states are assumed voltage independent).
Thus, in a simplified model the total qubit Hamiltonian is a sum of
a charging energy part (linear in gate voltages) and a (voltage independent) tunneling part,
see, e.g. Refs.~\cite{GroshevPRB1990,AverinKorotkovLikharevPRB1991,BeenakkerPRB1991}.
%
In a more elaborated QD models (see e.g. Ref.~\cite{DialYacoby2013PRL,Kerckhoff2020QST-HRL}) the qubit Hamiltonian
could acquire a voltage non-linearity.
Since, our derivations are general one, it is worth to mention other systems. E.g.
a Cooper pair box or transmon would possess a quadratic dependence on a gate voltage \cite{Devoret-AnnPhys:2007},
see Sec.~\ref{Sec-transmon-qubit} below.

One can now  expand a voltage dependent atom
Hamiltonian \cite{RuskovTahan2017preprint1,RuskovTahan2019PRB99,HarveyYacoby2018PRB,BottcherYacobyNatComm2022}
near a working point, $V_0$
\footnote{
Similar expansion was provided in Ref.~\cite{LevyYeyati2020PRL}.
}:
\begin{equation}
{\cal H}_{\rm qb}(V_0 + \delta V_G)
= {\cal H}_{\rm qb}(V_0) + \frac{\partial {\cal H}_{\rm qb}}{\partial V_G}\, \delta V_G
+ \frac{1}{2}\, \frac{\partial^2 {\cal H}_{\rm qb}}{\partial V_G^2} \, \delta V_G^2 ,
\label{Hamiltonian_expansion}
\end{equation}
where the gate voltage variation contains a quantum and classical part,
\begin{equation}
\delta V_G = v_q (\hat{a} + \hat{a}^{\dagger}) + \tilde{V}_m(t)
= v_q \, \left(\hat{a} + \hat{a}^{\dagger}
+ \bar{b}_{\rm cl} e^{-i\omega_m t} + \bar{b}_{\rm cl} e^{i\omega_m t}\right) ,
\end{equation}
$v_q \equiv \alpha_c V_{\rm vac}$ is the amplitude of quantum voltage fluctuations imposed on the QD system,
and $\bar{b}_{\rm cl} = \frac{\tilde{V}_m}{2 v_q}$, as in Eq.~(\ref{b_classical})
[$v_q$ is a small parameter and $\bar{b}_{\rm cl} \sim 1 \, {\rm or} \, \gg 1$].

The linear term in voltage variation leads to the usual static longitudinal and transverse couplings.
Indeed, the operator $\frac{\partial {\cal H}_{\rm qb}}{\partial V_G}$ has dimension of charge
and in the absence of modulation the atom-resonator interaction can be written
in the atom's eigenbasis, $\{|\psi_i\rangle \}$ as,
\begin{eqnarray}
&& \frac{\partial {\cal H}_{\rm qb}}{\partial V_G}\, \delta V_G
= \sum_{i,j} \langle \psi_i |\frac{\partial {\cal H}_{\rm qb}}{\partial V_G} | \psi_j\rangle \,
|\psi_i\rangle \langle \psi_j | \, v_q \, (\hat{a} + \hat{a}^\dagger)
\nonumber\\
&& \ \ = \sum_i \hbar g_{ii}\, |\psi_i\rangle \langle \psi_i |\, (\hat{a} + \hat{a}^\dagger)
+ \sum_{i\neq j} \hbar g_{ij}\, |\psi_i\rangle \,  \langle \psi_ j|\, (\hat{a} + \hat{a}^\dagger) , \qquad
\label{static_longitudinal_transverse}
\end{eqnarray}
where $\hbar g_{ij} \equiv v_q \langle \psi_i |\frac{\partial {\cal H}_{\rm qb}}{\partial V_G} | \psi_j\rangle$
plays the role of
a
dipole matrix element.
In Eq.~(\ref{static_longitudinal_transverse})  the single sum is the static longitudinal
interaction of the $n$-level atom.
Note, that using the Feynman-Hellman theorem,
$\langle \psi_i |\frac{\partial {\cal H}_{\rm qb}}{\partial V_G} | \psi_i\rangle
= \frac{\partial E_i}{\partial V_G}$.
Thus, it is reduced to the form presented in Ref.~\cite{RuskovTahan2019PRB99},
and the double sum (with $i\neq j$) is the usual transverse interaction.
In RWA, transverse interaction is reduced to the ``energy-conserving'' Jaynes-Cummings form,
while the static longitudinal interaction is suppressed.
Still, for two qubits, for example, the static longitudinal interaction  can lead
to qubits' entanglement, see, e.g. \cite{Billangeon2015PRB,RuskovTahan2021PRB103},
which amounts to interesting observable effects \cite{Billangeon2015PRB,RicherDiVincenzo2016PRB,SBoscoDLoss2022}.

Consider now the quadratic term in the Taylor expansion in Eq.~(\ref{Hamiltonian_expansion}),
which has the interpretation of atom's polarizability,
since $\delta^{(2)}{\cal H}_{\rm qb} \sim \tilde{\alpha}\, \delta V_G^2$
(also note that $\frac{\partial^2 {\cal H}_{\rm qb}}{\partial V_G^2}$ has dimension of capacitance).
Expanding in the atom's  eigenbasis  one obtains
\begin{eqnarray}
&& \frac{1}{2} \frac{\partial^2 {\cal H}_{\rm qb}}{\partial V_G^2}\, \delta V_G^2
= \sum_{i,j}
\langle \psi_i | \frac{1}{2} \frac{\partial^2 {\cal H}_{\rm qb}}{\partial V_G^2} | \psi_j\rangle \,\,
|\psi_i\rangle \langle \psi_j | \,\, \delta V_G^2
\nonumber\\
&& \qquad\quad \simeq \sum_i  \langle \psi_i | \frac{\partial^2 {\cal H}_{\rm qb}}{\partial V_G^2} | \psi_i\rangle \,
|\psi_i\rangle \langle \psi_i | \,
\nonumber\\
&& \qquad\quad  \times v_q^2 \left[\hat{a}^{\dagger} \hat{a} + \frac{1}{2}
+ \bar{b}_{\rm cl} \, \left(\hat{a} e^{-i \omega_m t} +  \hat{a}^{\dagger} e^{i \omega_m t} \right) \right] ,
\qquad\quad
\label{polarizability_effect}
\end{eqnarray}
where the last equation follows in the RWA.

For the dispersive interaction, $\sim | \psi_i\rangle \langle \psi_i| \, \hat{a}^{\dagger} \hat{a}$,
one combines the polarizability contribution from Eq.~(\ref{polarizability_effect}),
and the second order PT contribution from Eq.~(\ref{H-eff-1})
[both are of order $\sim v_q^2$] to obtain:
%
\begin{eqnarray}
&& {\cal H}_{\delta\omega} =
\sum_i | \psi_i\rangle \langle \psi_i| \, \hat{a}^{\dagger} \hat{a} \,
\left\{v_q^2 \langle \psi_i | \frac{\partial^2 {\cal H}_{\rm qb}}{\partial V_G^2} | \psi_i\rangle
+ D_i(\omega_r) \right\}
\nonumber\\
&& \qquad { } \equiv \sum_i  \delta\omega_i \ | \psi_i\rangle \langle \psi_i| \, \hat{a}^{\dagger} \hat{a} ,
\label{dispersive_Hamiltonian_general}
\end{eqnarray}
where $D_i(\omega_r)$ is given by the sum over dipole transitions of Eq.~(\ref{D_k})
and
\begin{equation}
\delta\omega_i \equiv  v_q^2 \langle \psi_i | \frac{\partial^2 {\cal H}_{\rm qb}}{\partial V_G^2} | \psi_i\rangle
+ D_i(\omega_r)
\label{dispersive_general_strength}
\end{equation}
is the effective dispersive interaction for the $i$-th level.

Combining the corresponding polarizability
and second order PT contributions, 
for the dynamical longitudinal interaction,
$\sim | \psi_i\rangle \langle \psi_i|  \left( \hat{a} + \hat{a}^{\dagger} \right) \cos(\omega_m t)$,
one obtains:
\begin{eqnarray}
&& {\cal H}_{\parallel} =
\sum_i  | \psi_i\rangle \langle \psi_i| \, (\hat{a} + \hat{a}^{\dagger} )\, \cos(\omega_m t) \,
\nonumber\\
&& \qquad\quad \times \left\{
v_q^2\, (2 \bar{b}_{\rm cl})\, \langle \psi_i | \frac{\partial^2 {\cal H}_{\rm qb}}{\partial V_G^2} | \psi_i\rangle
+ \tilde{L}_i(\omega_r,\omega_m) \right\}
\nonumber\\
&& \qquad { } \equiv \sum_i \tilde{g}_{\parallel,i}(\omega_r,\omega_m) \
| \psi_i\rangle \langle \psi_i| \, (\hat{a} + \hat{a}^{\dagger} )\, \cos(\omega_m t) ,
\qquad
\label{dynamical_longitudinal_Hamiltonian_general}
\end{eqnarray}
where $\tilde{L}_i(\omega_r,\omega_m) = \bar{b}_{\rm cl}\, \left[ D_i(\omega_r) + D_i(\omega_m) \right]$,
is given by Eq.~(\ref{L_k-modified}),
and
\begin{equation}
\tilde{g}_{\parallel,i}(\omega_r,\omega_m) =
v_q^2\, (2 \bar{b}_{\rm cl})\, \langle \psi_i | \frac{\partial^2 {\cal H}_{\rm qb}}{\partial V_G^2} | \psi_i\rangle
+ \tilde{L}_i(\omega_r,\omega_m)
\label{dynamical_longitudinal_strength}
\end{equation}
is the effective dynamical longitudinal coupling of the $i$-th level
in the general case of $\omega_m \neq \omega_r$.

We will use now a quantum-mechanical sum rule \cite{LevyYeyati2020PRL}
relating the polarizability matrix element and the energy curvature (quantum capacitance) of the level
via the dipole matrix elements (for completeness, it is re-derived in Appendix \ref{app-B: Low-energy-sum-rule}):
\begin{equation}
\langle \psi_i | \frac{\partial^2 {\cal H}_{\rm qb}}{\partial V_G^2} | \psi_i\rangle
= \frac{\partial^2 E_i}{\partial V_G^2}
+ 2 \sum_{j\neq i}
\frac{|\langle \psi_i | \frac{\partial {\cal H}_{\rm qb}}{\partial V_G} |\psi_j\rangle |^2}{E_j - E_i} .
\label{QuMech-sum-rule}
\end{equation}
The ``dispersive-like'' Hamiltonian, ${\cal H}_{\delta\omega}$, Eqs.~(\ref{dispersive_Hamiltonian_general})
and (\ref{dispersive_general_strength}) then obtains the dispersive couplings,
\begin{eqnarray}
&&\delta\omega_i = \left( v_q^2 \frac{\partial^2 E_i}{\partial V_G^2}
+ 2\sum_{j\neq i} \, \frac{|g_{ij}|^2}{\omega_{ji}} \right)
+ D_i(\omega_r)
\nonumber\\
&& { } = v_q^2 \frac{\partial^2 E_i}{\partial V_G^2}
+\sum_{j\neq i}|g_{ij}|^2 \,
\left( \frac{2}{\omega_{ji}} - \frac{1}{\omega_{ji} + \omega_r} - \frac{1}{\omega_{ji} - \omega_r}\right) ,
\qquad
\label{dispersive_general}
\end{eqnarray}
compare with Ref.~\cite{LevyYeyati2020PRL}.

In the resonant case, $\omega_m = \omega_r$,
the dynamical longitudinal Hamiltonian, ${\cal H}_{\parallel}$,
Eq.~(\ref{dynamical_longitudinal_Hamiltonian_general}), obtains the
couplings $\tilde{g}_{\parallel,i}$,
expressed  in a similar manner as a function of $\omega_r$:
\begin{equation}
\tilde{g}_{\parallel,i}(\omega_r) = \frac{\tilde{V}_m}{v_q}\,\, \delta\omega_i(\omega_r) ,
\label{dynamical_longitudinal_general}
\end{equation}
so that their ratio is independent of $\omega_r$.
Since typically, the voltage modulation can be made much larger than amplitude of
vacuum fluctuations, $\tilde{V}_m \gg v_q \equiv \alpha_c V_{\rm vac}$,
one gets the corresponding enhancement of the dynamical longitudinal coupling
vs. the dispersive coupling:
\begin{equation}
\tilde{g}_{\parallel,i}  \gg  \delta\omega_i .
\end{equation}

The dispersive Hamiltonian, ${\cal H}_{\delta\omega}$, Eqs.~(\ref{dispersive_Hamiltonian_general}),
and the dynamical longitudinal Hamiltonian, ${\cal H}_{\parallel}$,
Eq.~(\ref{dynamical_longitudinal_Hamiltonian_general}), with their partial interaction strengths $\delta\omega_i$,
$\tilde{g}_{\parallel,i}$
of Eqs.~(\ref{dispersive_general}) and (\ref{dynamical_longitudinal_strength})
[or Eq.~(\ref{dynamical_longitudinal_general})], 
respectively,
are the main results of this paper.

\section{Limiting regimes: adiabatic vs. dispersive}
\label{SecV-limiting-regimes}

\subsection{The adiabatic regime}
In the adiabatic limit when $\omega_r,\omega_m \ll |\omega_{kl}| \equiv |\omega_k - \omega_l|$ for any $k,l$,
and $\omega_r$, $\omega_m$, can be neglected in Eqs.~(\ref{dispersive_general}) and
(\ref{dynamical_longitudinal_strength}).
Then, {\it for any polarizability},
the corresponding strengths are expressed in terms of the energy curvature
of the levels as a result of the low-energy
QM  sum rule, Eq.~(\ref{QuMech-sum-rule}):
\begin{eqnarray}
&& \delta\omega_i(\omega_r=0) = v_q^2 \frac{\partial^2 E_i}{\partial V_G^2}
\label{adiabatic-limit-delta-omega}
\\
&&  \tilde{g}_{\parallel,i}(\omega_r=0,\omega_m=0) = v_q \tilde{V}_m \frac{\partial^2 E_i}{\partial V_G^2} .
\label{adiabatic-limit-g}
\end{eqnarray}
Thus, we recover the $n$-level adiabatic expressions,
Eqs.~(\ref{qudit-dispersive-like}) and (\ref{qudit-dynamical-longitudinal}),
derived via
Taylor expansion of small and slow voltage variations
as in Refs.~\cite{RuskovTahan2017preprint1,RuskovTahan2019PRB99}.

\subsection{QD spin-qubits: zero polarizability}

In the case of QD spin-qubits the qubit Hamiltonian, ${\cal H}_{\rm qb}^{\rm QDs}$ is linear
in the applied QDs' gate voltages, $V_{G,i}$ (in a simple model when interdot tunnelings $t_i$
are not affected by $V_{G,i}$).
Then, the polarizability matrix element is zero in Eqs.~(\ref{QuMech-sum-rule}), (\ref{dispersive_general}),
\begin{equation}
v_q^2 \,\, \langle \psi_i | \frac{\partial^2 {\cal H}_{\rm qb}}{\partial V_G^2} | \psi_i\rangle
\equiv v_q^2 \frac{\partial^2 E_i}{\partial V_G^2}
+ 2\sum_{j\neq i} \, \frac{|g_{ij}|^2}{\omega_{ji}} = 0
\label{zero-polarizability}
\end{equation}
and one ends up  with a simplified expression, Eq.~(\ref{D_k}), for
the effective dispersive coupling $\delta\omega_i$:
\begin{equation}
\delta\omega_i =
- \sum_{j\neq i} \, |g_{ij}|^2 \,
\left(\frac{1}{\omega_{ji} + \omega_r} + \frac{1}{\omega_{ji} - \omega_r}\right) \equiv D_i(\omega_r),
\label{D_k1}
\end{equation}
which coincides with the result of Ref.~\cite{Zhu:2013}.
As per
Eq.~(\ref{dynamical_longitudinal_general}),
the dynamical longitudinal couplings
in the resonant case
are simplified to:
\begin{equation}
\tilde{g}_{\parallel,i} = L_i(\omega_r) \equiv \frac{\tilde{V}_m}{\alpha_c V_{\rm vac}}\, D_i(\omega_r) .
\end{equation}

\subsection{Dispersive regime for QD spin-qubits}
By definition, in the dispersive regime the resonator frequency is of the order of a particular
qudit's energy difference,
$\omega_r \approx |\omega_l - \omega_{l'}|$, while the corresponding detuning
is large,
$|\omega_l - \omega_{l'} - \omega_r| \gg g_{ll'}$ for any $l,l'$.
Then, the effective Hamiltonian, ${\cal H}_{\delta\omega}$ recovers the well known dispersive interaction.

For example, in a two level approximation, and neglecting a small term,
$\sim \frac{g_{12}^2}{\omega_q + \omega_r} \ll \frac{g_{12}^2}{\omega_q - \omega_r}$,
one obtains from Eq.~(\ref{D_k1}),
\begin{equation}
{\cal H}_{\rm disp}/\hbar
= \frac{g_{12}^2}{\omega_q - \omega_r}\, \sigma_z\, \hat{a}^\dagger \hat{a}
\equiv \delta\omega^{\rm disp}\, \sigma_z\, \hat{a}^\dagger \hat{a} ,
\label{dispersive-interaction-limit}
\end{equation}
where
$\omega_q \equiv \omega_2 - \omega_1$ and
$\delta\omega^{\rm disp} \equiv \chi = \frac{g_{12}^2}{\omega_q - \omega_r}$.
It should be noted that in the adiabatic regime, which is formally similar to the dispersive regime
(with $\Delta = \omega_q$),
one should keep both terms in Eq.~(\ref{D_k1}) which are exactly equal to each other.
Then,
${\cal H}_{\delta\omega}/\hbar \approx
2 \frac{g_{12}^2}{\omega_q}\, \sigma_z\, \hat{a}^\dagger \hat{a}$, i.e.,
the dispersive coupling in the adiabatic regime is twice the standard dispersive coupling
of
the dispersive regime.

By considering the dynamical longitudinal coupling in the dispersive limit,
one obtains
in a two-level approximation:
\begin{equation}
{\cal H}_{\parallel,\,\rm disp}/\hbar
= \frac{e \tilde{V}_m}{2\hbar g_0}\, \frac{g_{12}^2}{\omega_q - \omega_r}\, \sigma_z\, (\hat{a}^\dagger + \hat{a})\,
\cos(\omega_m t)
\label{dispersive-longitudinal-limit} .
\end{equation}
The enhancement of both couplings going from the adiabatic regime to the dispersive regime
can be estimated as (assuming $|\omega_q - \omega_r|\approx 10 g_{12}$):
\begin{equation}
\frac{\tilde{g}_{\parallel}^{\rm disp}}{\tilde{g}_{\parallel}^{\rm adiabat}}
= \frac{\delta\omega^{\rm disp}}{\delta\omega^{\rm adiabat}} \approx \frac{\omega_q}{20 g_{12}}
\approx 12.5 - 50,
\label{enhancement_factor}
\end{equation}
for $\omega_q = 10\, {\rm GHz}$ and $g_{12} \approx 10 - 40\, {\rm MHz}$.

\section{Illustrative examples of qubits}
\label{SecVI-examples}

\subsection{DQD 1e charge qubit}
\label{Sec-charge-qubit}

For a charge qubit one considers a double quantum dot (DQD) with a single electron.
The Hamiltonian in the charge basis
of left (dot 1) and right (dot 2) localized states is
(scf. Fig.~\ref{fig:1}):
\begin{equation}
{\cal H}_q = \frac{\varepsilon}{2} \tilde{\sigma}_z + t_c \tilde{\sigma}_x
\label{dqd-1e-model} ,
\end{equation}
where $\varepsilon = e (V_1 - V_2)$ is the double dot energy detuning parameter,
$V_1$, $V_2$ are the gate voltages applied to the dots,
$t_c$ is the tunneling,
and
$\tilde{\sigma}_z$, $\tilde{\sigma}_x$ are the Pauli matrices in this basis.
The qubit-resonator dipole coupling ($\vec{d}\cdot \vec{E}$) arises
via the resonator quantized voltage, $\hat{V}_r$, see Eq.~(\ref{vac_field}).

The resonator
is coupled to dot 2 so that the quantum voltage change,
$\delta \hat{V}_2 = \alpha_c \hat{V}_r$, at the dot 2 is given by the lever arm,
$\alpha_c \simeq \frac{C_c}{C_c + C_d}$,
where $C_c$ is the dot 2 to resonator capacitance, and $C_d$ is the dot to ground capacitance.
The corresponding energy change is given by
\begin{equation}
\delta {\cal H}_q = \frac{\partial {\cal H}_q}{\partial V_2} \delta \hat{V}_2
= - \frac{1}{2} (e\, \tilde{\sigma}_z) (\alpha_c \hat{V}_r)
\label{DQD-energy-change} ,
\end{equation}
and results in a DQD-resonator dipole interaction:
\begin{equation}
\delta {\cal H}_q = \hbar g_0 \tilde{\sigma}_z (a + a^\dagger)
\label{DQD-dipole-interaction} .
\end{equation}

By diagonalizing the qubit Hamiltonian ${\cal H}_q$ at a fixed detuning $\varepsilon$,
one obtains the total DQD plus resonator Hamiltonian:
\begin{equation}
{\cal H}_{\rm tot}/\hbar = \frac{\omega_q}{2} \sigma_z + \omega_r a^\dagger a +
\left(g_{\parallel}^{\rm static} \sigma_z + g_{\perp} \sigma_x \right) (a + a^\dagger)
\label{total-H-0} ,
\end{equation}
where $\sigma_z$, $\sigma_x$, are the Pauli matrices in the qubit eigenbasis,
$|+\rangle$, $|-\rangle$, with energies,
$E_{q0,\pm} = \pm \frac{E_q}{2}$
with $E_q = \hbar \omega_q = \sqrt{\varepsilon^2 + 4 t_c^2}$.
In Eq.~(\ref{total-H-0})
the $g_{\parallel}^{\rm static}$ and $g_{\perp}$ are the static longitudinal and transverse coupling,
respectively:
\begin{eqnarray}
&&  g_{\parallel}^{\rm static} =
g_0\,
\frac{\varepsilon}{\sqrt{\varepsilon^2 + 4 t_c^2} }, \ \ \
g_{\perp} =
g_0\,
\frac{2 t_c}{\sqrt{\varepsilon^2 + 4 t_c^2} }
\label{longitude-transverse-dipole-couplings} ,
\end{eqnarray}
that correspond to diagonal ($\sim g_{\parallel}^{\rm static}$) and
transition ($\sim g_{\perp}$) dipole matrix elements
of Eq.~(\ref{static_longitudinal_transverse}).
In a rotating-wave approximation 
the former is suppressed while the latter
obtains the energy-conserving Jaynes-Cummings form
$\propto g_{\perp} \left(a \sigma_{+} + a^\dagger \sigma_{-} \right)$.

In the adiabatic limit one can obtain the effective curvature (quantum capacitance) couplings
from the ``soft-field'' formulae, Eqs.(\ref{qudit-dispersive-like}) and (\ref{qudit-dynamical-longitudinal})
of Sec.~\ref{SecIII-adiabatic},
using the energy level dependence of $E_q(\varepsilon)$.
One then obtains the ``dispersive-like''
and dynamical longitudinal (curvature) Hamiltonians of
Eqs.~(\ref{qubit_curvature_Hamiltonian-disp-like}) and (\ref{qubit_curvature_Hamiltonian-lonitudinal}),
with strengths:
\begin{eqnarray}
&& \delta\omega
= 2\hbar g_0^2\, \frac{\partial^2 E_q}{\partial \varepsilon^2}
= \hbar g_0^2  \,
\frac{8 t_c^2}{[\varepsilon^2 + 4 t_c^2]^{3/2}}
\label{dispersive-like-coupling} .
\end{eqnarray}
and
\begin{eqnarray}
&& \tilde{g}_{\parallel}
= g_0\, \frac{\partial^2 E_q}{\partial \varepsilon^2} e\,\tilde{V}_{m}
= g_0 \, \frac{4 t_c^2}{[\varepsilon^2 + 4 t_c^2]^{3/2}} e \tilde{V}_{m} \quad\
\label{dynamic-long-coupling-parallel} .
\end{eqnarray}
(same result can be obtained by modulating Eq.~(\ref{total-H-0}),
compare with Ref.~\cite{Didier2015PRL}).
The above dynamical longitudinal coupling $\tilde{g}_{\parallel}$ is on/off together with the
dot's gate voltage modulation.

By using the perturbative method of Sec.~\ref{SecIV-A-perturbative} one can obtain
the more general effective couplings that depend on the resonator frequency.
By taking into account the zero polarizability condition for QDs Hamiltonians, Eq.~(\ref{zero-polarizability}),
one obtains for a charge qubit system (in a two-level approximation)
the effective couplings,
\begin{equation}
\delta\omega_{\rm eff} = g_{\perp}^2\,
\left(\frac{1}{\omega_q - \omega_r} + \frac{1}{\omega_q + \omega_r} \right)
\label{delta-omega-eff}  ,
\end{equation}
while $\tilde{g}_{\parallel, \rm eff} = \frac{\tilde{V}_m}{v_q}\, \delta\omega_{\rm eff}$.
In Eq.~(\ref{delta-omega-eff}),
in the limit $\omega_r \ll \omega_q$,
one obtains $\delta\omega_{\rm eff} \approx 2 \frac{g_{\perp}^2}{\omega_q}$, which coincides
with the ``soft-field'' expression, Eq.~(\ref{dispersive-like-coupling}),
as has been shown in general in Sec.~\ref{SecV-limiting-regimes}.

We note that recently,
to reveal the curvature couplings in
adiabatic limit,
an experiment was performed on a hybrid spin-qubit \cite{CorriganHarptRuskovEriksson2023PRAppl}
(approximated as
a  two-level charge qubit),
showing a clear signature of the adiabatic longitudinal interactions,
Eqs~(\ref{dispersive-like-coupling}) and (\ref{dynamic-long-coupling-parallel}).

\subsection{The transmon qubit}
\label{Sec-transmon-qubit}

The transmon is the capacitevely shunted (by $C_S$) Josephson junction with energy $E_J$.
In the charge basis (of states with definite number (N) of Cooper pairs (2e)
on the transmon island one has
the Hamiltonian:
\begin{eqnarray}
&& {\cal H}_{\rm tran} = 4 E_C \sum_N (N - N_g)^2 |N\rangle\langle N |
\nonumber\\
&& \qquad\qquad { } - \frac{E_J}{2}\, \left( |N\rangle\langle N+1 | + |N +1 \rangle\langle N | \right)
\label{transmon}
\end{eqnarray}
where $N_g = \frac{C_g V_g}{2 e}$ is the induced charge on the island by
a capacitively coupled voltage source $V_g$, $E_C \simeq \frac{(2 e)^2}{2 C_S}$ is the
island charging energy scale ($C_S \gg C_J, C_g$),
and $E_J$ plays the role of charge tunneling matrix element.

The derivatives of ${\cal H}_{\rm tran}$ are defined
with respect to the voltage at the island, $\tilde{V}_g = \frac{C_g}{C_g + C_S} \, V_g \equiv \alpha_g\, V_g$,
and the matrix element of the first derivative defines the corresponding transverse
coupling, $g_{ij}$, see Eq.~(\ref{static_longitudinal_transverse}):
\begin{equation}
\hbar g_{ij} = \alpha_g V_{\rm vac}\,
\langle \psi_i | \frac{\partial {\cal H}_{\rm tran}}{\partial \tilde{V}_g} |\psi_j \rangle
= - 2\alpha_g (e V_{\rm vac})\, \langle \psi_i | \hat{N} |\psi_j \rangle ,
\label{transverse_coupling_g_ij}
\end{equation}
see Ref.~\cite{JKoch:2007}.
The polarizability contribution to the $i$-th level dispersive shift
in Eq.~(\ref{dispersive_general_strength}) is then obtained
\begin{equation}
\delta\omega_i^P = v_q^2\,
\langle \psi_i | \frac{\partial^2 {\cal H}_{\rm tran}}{\partial \tilde{V}_g^2} |\psi_i \rangle
= \alpha_g^2 V_{\rm vac}^2 (C_S + C_g) = \frac{C_g^2}{C_S + C_g}\, V_{\rm vac}^2 ,
\label{polarizability_delta_omega}
\end{equation}
that is independent of the energy level.

\begin{figure} [t!] 
    \centering
     \includegraphics[width=0.46\textwidth]{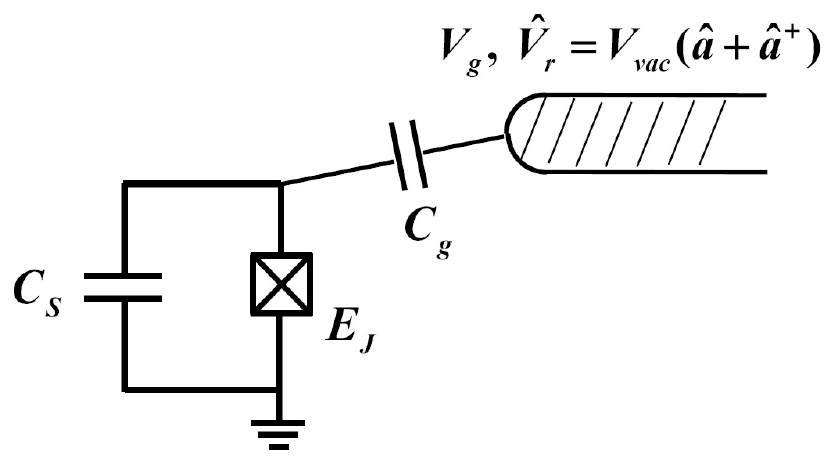}
        \caption{
        A schematic of transmon, capacitively coupled via $C_g$ to a voltage source, $V_g$ (or $\hat{V}_r$)
        (compare with Ref.~\cite{JKoch:2007}).
        $C_S$ is the shunting capacitance of the Josephson junction with energy $E_J$.
        }
        \label{fig:3}
\end{figure}
Since in the transmon limit, $E_J \gg E_C$, the curvature of the levels is zero,
$\frac{\partial^2 E_i}{\partial \tilde{V}_g^2} \simeq 0$, then the polarizability contribution,
is entirely due to the sum of
transition
dipole matrix elements contributions in Eq.~(\ref{QuMech-sum-rule}).
It is convenient to denote these matrix elements as
$m_{ij} \equiv \langle \psi_i | \frac{\partial {\cal H}_{\rm tran}}{\partial \tilde{V}_g} |\psi_j \rangle$.
By using the  transmon energy levels,
$E_i \simeq -E_J + \sqrt{8 E_C E_J} - \frac{E_C}{12}\, (6 i^2 + 6 i +3)$,
and that for given `i' only the nearest neighbor dipole matrix elements survive in the transmon limit \cite{JKoch:2007},
$|\langle i+1 | \hat{N} |i \rangle| \simeq \sqrt{\frac{i+1}{2}}\, \left(\frac{E_J}{8 E_C} \right)^{1/4}$,
one can show that for any energy level `i' the r.h.s. of Eq.~(\ref{QuMech-sum-rule}) provides
the same expression as the direct second order derivative,
$\langle \psi_i | \frac{\partial^2 {\cal H}_{\rm tran}}{\partial \tilde{V}_g^2} |\psi_i \rangle$,
scf. Eq.~(\ref{polarizability_delta_omega}), as expected:
\begin{eqnarray}
&&\frac{\partial^2 E_i}{\partial \tilde{V}_g^2} + 2 \sum_{j\neq i}\, \frac{|m_{ij}|^2}{E_j - E_i}
\nonumber\\
&& \simeq \ldots + 2 \frac{|m_{i-1,i}|^2}{E_{i-1} - E_i}
+ 2 \frac{|m_{i+1,i}|^2}{E_{i+1} - E_i} + \ldots
\nonumber\\
&& \simeq 4e^2 \left( \frac{E_J}{8 E_C} \right)^{1/2} \frac{1}{\sqrt{8 E_C E_J}} = C_S + C_g ,
\label{transmon-sum}
\end{eqnarray}
where the second and third row are written in the transmon limit.

While in the adiabatic limit, $\omega_r \ll |\omega_{ll'}|$, all couplings
are zero
(in the transmon limit the energy levels are flat),
$\delta\omega_i = 0$, $\tilde{g}_{\parallel} = 0$,
in the dispersive regime
one recovers the known expressions for the dispersive coupling (restricting to three levels;
see, e.g. Ref.~\onlinecite{JKoch:2007}):
\begin{eqnarray}
&& \delta\omega_0^{\rm disp} \approx - \frac{|g_{01}|^2}{\omega_{10} - \omega_r} \equiv -\chi_{01}
\\
&& \delta\omega_1^{\rm disp} \approx \frac{|g_{01}|^2}{\omega_{10} - \omega_r} - \frac{|g_{12}|^2}{\omega_{21} - \omega_r}
\equiv \chi_{01}-\chi_{12}
\label{transmon_dispersive_shifts}
\\
&& \chi_{\rm eff}^{\rm disp} \approx \frac{\delta\omega_1^{\rm disp} - \delta\omega_0^{\rm disp}}{2}
= \chi_{10} - \frac{1}{2}\chi_{12} ,
\label{transmon_chi_eff}
\\
&& {\cal H}_{\rm disp} = \chi_{\rm eff}^{\rm disp} \sigma_z a^\dagger a .
\qquad
\label{effective_dispersive_delta_omega}
\end{eqnarray}
For the dynamical longitudinal coupling in the dispersive regime, one then obtains
\begin{eqnarray}
&& \tilde{g}_{\parallel,\rm eff}^{\rm disp}
= \frac{\tilde{V}_m}{\alpha_g V_{\rm vac}}\, \left(\chi_{10} - \frac{1}{2}\chi_{12}\right)
\label{dynamical_longitudinal_eff}
\\
&& {\cal H}_{\parallel, \rm disp}
= \tilde{g}_{\parallel,\rm eff}^{\rm disp}\ \, \sigma_z\, (a + a^\dagger)\, \cos(\omega_r t)
\label{effective_dispersive_dynamical_longitudinal}
\end{eqnarray}
We note that the first term in Eq.~(\ref{dynamical_longitudinal_eff})
coincides with a similar longitudinal coupling derived in a different approach
in Ref.~\cite{Touzard:2019}.

When one deviates from the dispersive regime (by considering smaller $\omega_r$)
one obtains the full expression for $\chi_{\rm eff}$
(in the transmon limit),
which interpolates between $\chi_{\rm eff}^{\rm disp}$ of Eq.~(\ref{transmon_chi_eff})
and zero
\begin{eqnarray}
&& \chi_{\rm eff} \simeq \frac{\delta\omega_1 - \delta\omega_0}{2}
= - |g_{01}|^2\, \left(\frac{2}{\omega_{10}}
- \frac{1}{\omega_{10} + \omega_r} - \frac{1}{\omega_{10} - \omega_r} \right)
\nonumber\\
&& \qquad { } + \frac{1}{2}\, |g_{12}|^2\, \left(\frac{2}{\omega_{21}}
- \frac{1}{\omega_{21} + \omega_r} - \frac{1}{\omega_{21} - \omega_r} \right) .
\end{eqnarray}

Out of the transmon limit one has to take into account
the whole sum over transition dipole matrix elements
in Eqs.~(\ref{dispersive_general}) and (\ref{dynamical_longitudinal_general}),
see also Eq.~(\ref{transmon-sum}),
which we did not show here explicitly.
The expression for  the dynamical longitudinal coupling
has the same functional dependence since the proportionality of the two couplings,
$\tilde{g}_{\parallel, \rm eff} = \frac{\tilde{V}_m}{\alpha_g V_{\rm vac}}\, \chi_{\rm eff}$
(in the resonance case).

One notes that for large detunings, when $\Delta_{10}=\omega_{10} - \omega_r$ becomes comparable
to the transmon splitting, the relative difference,
$\frac{\chi_{\rm eff} - \chi_{\rm eff}^{\rm disp}}{\chi_{\rm eff}^{\rm disp}}$ can reach $\lesssim \% 100$.


\subsection{DQD Singlet-Triplet qubit}
\label{Sec-DQD-ST-qubit}

Consider now a DQD Singlet-Triplet (S-T) system in which the qubit states
are given mainly by the charge configurations $(1,1)$ with some admixture of the
higher charge states \cite{RuskovTahan2019PRB99}, $|S(0,2)\rangle$ and $|S(2,0)\rangle$,
i.e., the
qubit
ground state $|-\rangle \approx |S(1,1)\rangle$ and the excited state
$|+\rangle \approx |T_0(1,1)\rangle$, both having zero spin projection, $S_z=0$.
Due to tunneling (with amplitude $t_c$) between singlet charge configurations, the ground state $|-\rangle$
obtains curvature as a function of the DQD energy detuning, $\varepsilon = e(V_1 - V_2)$
(compare with the charge qubit, Sec.~\ref{Sec-charge-qubit}).
The excited   
state, $|+\rangle$,
remains flat, essentially, due to Pauli spin blockade.

In what follows, we will consider detuning regimes when one of the charge states (e.g., $|S(2,0)\rangle$)
is highly gapped by the dot's charging energy ($E_{\rm gap} \sim U_{\rm charge} \sim {\rm hundreds\ GHz}$),
and can be neglected.
In the remaining three-level system, the upper charge state $|S(0,2)\rangle$ is gapped from the qubit
ground state, $|-\rangle$,  
by several tens of GHz, reaching a minimum of $2 t_c$ at the charge degeneracy point (c.d.p.).
%
By measuring the detuning $\varepsilon $ from the c.d.p.
the three-level Hamiltonian in the charge basis,
$\{T_0(1,1), S(1,1), S(0,2)\}$, reads:
\begin{equation}
{\cal H}_{\rm DQD}^{\rm c.d.p.} =
\left(
\begin{array}{ccc}
0  & 0             & 0
\\
0  & 0             &  t_c
\\
0  & t_c  &  -\varepsilon   
\end{array}
\right)
\label{DQD-Hamiltonian-cdp}  .
\end{equation}
For the eigenergies one gets
\begin{eqnarray}
&& E_{-}(\varepsilon) = - \frac{\varepsilon}{2} - \frac{1}{2}\sqrt{\varepsilon^2 + 4 t_c^2}
\\
&& E_{+}(\varepsilon) = 0
\\
&& E_{\rm S(0,2)}(\varepsilon) = - \frac{\varepsilon}{2} + \frac{1}{2}\sqrt{\varepsilon^2 + 4 t_c^2} .
\label{S-T-qb-energy-levels}
\end{eqnarray}
By calculating the curvatures of the levels one obtains
from Eqs.~(\ref{qudit-dispersive-like}) and (\ref{qudit-dynamical-longitudinal})
the effective ``dispersive-like'' and dynamical longitudinal interactions
(projected on the qubit space):
\begin{eqnarray}
&& \delta\omega = 2\hbar g_0^2 \frac{\partial^2 E_q}{\partial \varepsilon^2},\ \ \
\tilde{g}_{\parallel} = \frac{e\tilde{V}_m}{2\hbar g_0}\, \delta\omega
\\
&&  \frac{\partial^2 E_q}{\partial \varepsilon^2}
= \frac{1}{2} \frac{4 t_c^2}{[\varepsilon^2 + 4 t_c^2]^{3/2}} .
\end{eqnarray}
The qubit energy curvature is 1/2 of that for a charge qubit, since the contribution from the
(``flat'') state $|T_0(1,1)\rangle$ is zero.

To obtain the effective interactions to the resonator in a non-adiabatic regime
one needs the dipole interactions of the levels.
Similar to the charge qubit case above, one writes the dipole interaction
using the linear response approach \cite{RuskovTahan2019PRB99}.
In the system's charge basis the dipole interaction is diagonal:
\begin{equation}
{\cal H}_{\rm DQD,\ dipole}^{\rm c.d.p.} \simeq
2 \hbar g_0\, (a + a^\dagger)
\left(
\begin{array}{ccc}
1  & 0             & 0
\\
0  & 1             & 0
\\
0  & 0  &  2
\end{array}
\right)
\label{DQD-cdp-dipole}  .
\end{equation}
In the eigenbasis one then obtains a diagonal dipole part
(corresponding to a static longitudinal interaction)
\begin{equation}
{\cal H}_{\rm DQD, \parallel}^{\rm stat} \simeq
2 \hbar g_0 \, (a + a^\dagger)
\left(
\begin{array}{ccc}
1  & 0             & 0
\\
0  & \frac{3}{2} + \frac{\varepsilon}{2\sqrt{\varepsilon^2 + 4t^2}}     & 0
\\
0  & 0  &  \frac{3}{2} - \frac{\varepsilon}{2\sqrt{\varepsilon^2 + 4t^2}}
\end{array}
\right)
\label{DQD-cdp-dipole-stat-longitudinal}  ,
\end{equation}
and transverse dipole part
\begin{equation}
{\cal H}_{\rm DQD, \perp} \simeq
2 \hbar g_0 \, (a + a^\dagger)
\left(
\begin{array}{ccc}
0  & 0             & 0
\\
0  & 0             & -\frac{t_c}{\sqrt{\varepsilon^2 + 4t^2} }
\\
0  & -\frac{t_c}{\sqrt{\varepsilon^2 + 4t^2} }  &  0
\end{array}
\right)
\label{DQD-cdp-dipole-transverse}  ,
\end{equation}
according to the general structure of Eq.~(\ref{static_longitudinal_transverse}).
We would stress the following points.
First, while in a RWA the (static) longitudinal dipole part is suppressed,
a time modulation of the detuning
survives
in RWA and
will lead exactly to the expression for a qudit dynamical
longitudinal coupling Hamiltonian of Eq.~(\ref{qudit-dynamical-longitudinal})
(for a specially designed superconducting qubit such interaction was derived in Ref.~\cite{Didier2015PRL}).

More important for our further study is the transverse dipole part
(here and below we use the enumeration of the states:
$|0\rangle \equiv |-\rangle \approx |S(1,1)\rangle$, $|1\rangle\equiv |+\rangle \approx |T_0(1,1)\rangle$,
$|2\rangle \equiv|S(0,2)\rangle$):
A non-zero transition dipole matrix element exists only between the states
$|-\rangle$ and $|S(0,2)\rangle$,
\begin{equation}
g_{02} = -2 g_0\, \frac{t_c}{\sqrt{\varepsilon^2 + 4t^2} }
\label{S-T-dipole-matrix-element} ,
\end{equation}
while the other two dipole couplings
are zero, $g_{21}=0$, $g_{01}=0$. This selection rule is essentially due to Pauli spin blockade.
With these dipole matrix elements it is straightforward to show
that the effective dispersive/longitudinal couplings of the S-T qubit are
\begin{eqnarray}
&&\delta\omega_{\rm eff} = \frac{|g_{02}|^2}{2} \,
\left( \frac{1}{\omega_{20} + \omega_r} + \frac{1}{\omega_{20} - \omega_r}\right)
\label{delta-omega-eff-ST}
\\
&& \tilde{g}_{\parallel, \rm eff} = \frac{e \tilde{V}_m}{2\hbar g_0}\, \delta\omega_{\rm eff}
\end{eqnarray}
where we take into account that for QDs' systems the polarizability contribution,
Eq.~(\ref{QuMech-sum-rule}), is zero, see Sec.~\ref{Sec-charge-qubit}.
For the adiabatic limit we take $\omega_r \ll \omega_{20}$, since only the states
$|0\rangle$, $|2\rangle$, are dipole coupled.
Then,
\begin{equation}
\delta\omega_{\rm eff}^{\rm adiabat} \simeq \frac{|g_{02}|^2}{\omega_{20}}
= 2\hbar g_0^2\, \frac{\partial^2 E_q}{\partial \varepsilon^2} ,
\label{delta_omega_adiabatic}
\end{equation}
where the last equation reproduces the general low-energy sum rule,
Eqs.~(\ref{adiabatic-limit-delta-omega}) and (\ref{adiabatic-limit-g}).
We note that    
$\delta\omega_{\rm eff}$
is $1/2$ of the expression
for a charge qubit, Eq.~(\ref{delta-omega-eff}),
since only the curvature of the ground state contributes.

While the adiabatic limit implies $\omega_r \ll \omega_{20}$,
the resonator frequency can be comparable or even larger than the qubit frequency, $\omega_{10}$.
So, the dispersive regime in Eq.~(\ref{delta-omega-eff-ST})
would correspond to $\omega_r \approx \omega_{20}$ and $|\omega_{20} - \omega_r| \gg g_{02}$.

From these considerations, and from the general expressions for the quantum measurement
rate $\Gamma_{\rm meas}(\tilde{g}_{\parallel, \rm eff},\delta\omega_{\rm eff})$,
considered in Ref.~\cite{RuskovTahan2019PRB99}, it follows that the
general strategy to perform a S-T qubit
strong quantum measurement is to be in the dispersive regime
with respect to the third level, $|S(0,2)\rangle$,
and to use the enhancement of the coupling strength via the dynamical longitudinal coupling.

\subsection{TQD exchange only qubit}
\label{Sec-TQD exchange only qubit}

Similar analysis can be performed for the TQD exchange only qubit
\cite{RuskovTahan2019PRB99}. 
At and around the symmetric operating point \cite{AEON2016,RuskovTahan2019PRB99} (SOP is a double sweet spot)
the sum over dipole matrix elements for
the ``dispersive-like'' and dynamical longitudinal couplings, $\delta\omega$ and $\tilde{g}_{\parallel}$,
of Eqs.~(\ref{dispersive_general_strength}) and (\ref{dynamical_longitudinal_strength})
are dominated by the dipole transitions
from the qubit states, $|-\rangle$, $|+\rangle$
[of charge configuration $(1,1,1)$],
to the 4 highly gapped ($E_{\rm gap} \approx U_{\rm charge}$) charge states, all of spin $S_z = 1/2$,
i.e.,
$|3\rangle \equiv (2,0,1) \equiv |S(2_1,0_2)\, \uparrow_3 \rangle$,
$|4\rangle \equiv (1,0,2)\equiv |\uparrow_1 \, S(0_2,2_3)\rangle$,
and similar for
$|5\rangle \equiv (1,2,0)$,
$|6\rangle \equiv (1,0,2)$.
These dipole couplings are denoted as
$g_{-,l}$ and $g_{+,l}$, $l=3,4,5,6$ and calculated in Appendix \ref{app-C: Dipole-m-e-TQD-qubit},
using linear response approach \cite{RuskovTahan2019PRB99}.
At (out of) the SOP the qubit dipole element, $g_{-,+}$ is
of zero (small non-zero)
value \cite{RuskovTahan2019PRB99}.
[For definitions of the states, dipole elements, and qubit curvature
see, Appendix \ref{app-C: Dipole-m-e-TQD-qubit} and Ref.~\cite{RuskovTahan2019PRB99}].

In a two-level approximation in the adiabatic limit, $\omega_r \ll E_{\rm gap} \sim U_{\rm charge}$,
(well fulfilled here since $\omega_r \lesssim 10\, {\rm GHz}$ and $U_{\rm charge} \gtrsim 100\, {\rm GHz}$)
%
\begin{equation}
\delta\omega_{\rm eff}^{\rm adiabat} = 2 \frac{|g_{-,+}|^2}{\omega_{+,-}}
+ \sum_{l=3}^6 \, \left( \frac{|g_{-,l}|^2}{\omega_{l,-}} - \frac{|g_{+,l}|^2}{\omega_{l,+}} \right)
\end{equation}
Using the dipole matrix elements, $g_{-,l}$ and $g_{+,l}$,
one can recover the curvature (quantum capacitance) couplings,
that is the dynamical longitudinal and ``dispersive-like'' \cite{RuskovTahan2019PRB99},
$\tilde{g}_{\parallel, \rm eff}^{\rm adiabat}, \delta\omega_{\rm eff}^{\rm adiabat} \propto \frac{\partial^2 E_q}{\partial V_m^2}$,
in the adiabatic limit, Eq.~(\ref{adiabatic-limit-delta-omega}).

It should be noted that at the SOP a dispersive regime,
when $\omega_r \sim U_{\rm charge}$, is not reachable.
Out of SOP, but still in the deep $(1,1,1)$ region, the dispersive regime would correspond
to $\omega_r \sim \omega_{+,-}$.
In this case the qubit levels dispersive coupling, $\chi = \frac{|g_{-,+}|^2}{\omega_{+,-} - \omega_r}$
{\it may compete}
with the ``dispersive-like'' coupling $\delta\omega_{\rm higher\ levels}$ induced by the
higher levels \cite{RuskovTahan2019PRB99}.

\subsection{Implications for quantum measurements}

Below we consider implications for reaching quantum limited regime
of continuous quantum measurements
on the promising example of a spin-charge qubit \cite{MiPetta2018N}.
While the spin-charge qubit is essentially a $\Lambda$-system
\cite{Benito2017PRB}
(with levels 0, 1, 2),
the relevant dynamical longitudinal coupling $\tilde{g}_{\parallel}(0,1,2)$
can be made large with respect to the dispersive coupling $\delta\omega(0,1,2)$,
by the possibly large ratio,
Eq.~(\ref{dynamical_longitudinal_general}), since
$\frac{\tilde{g}_{\parallel}(\omega_r)}{\delta\omega(\omega_r)}
= \frac{\tilde{V}_m}{\alpha_c V_{\rm vac}} \simeq 10 - 100$.
For (continuous) quantum measurements
in general, both curvature Hamiltonians, ${\cal H}_{\delta\omega}$ and ${\cal H}_{\parallel}$,
contribute to the measurement rate \cite{RuskovTahan2017preprint1,RuskovTahan2019PRB99}:
$\Gamma_{\rm meas} = \Gamma_{\delta\omega} + \Gamma_{\parallel}$.
However, in what follows, in the estimations we  will focus on the dynamical longitudinal
coupling only,
since it can dominate the measurement rate.

Indeed, in the so-called ``bad cavity limit'' (see, e.g. Refs.~\cite{Korotkov2011-Les_Houches,Korotkov2016PRA94})
one requires that
the
resonator damping rate is much faster, $\kappa \gg \Gamma_{\rm meas}$,
so that one is measuring the qubit alone (and not the combined system
of
qubit plus resonator),
see, e.g. Ref.~\cite{KorotkovSiddiqi2014PRL}.
Thus, one has the conditions, $\kappa^2 \gg 2\,\tilde{g}_{\parallel}^2, \, \delta\omega^2$,
and therefore
\begin{equation}
\frac{\Gamma_{\parallel}}{\Gamma_{\delta\omega}}
\approx \frac{\tilde{g}_{\parallel}^2}{4 \delta\omega^2} \,
\frac{\kappa^2/4}{\left| \varepsilon_d^2 - \tilde{g}_{\parallel}^2/4 \right|} \gg 1
\label{ratio-partial-rates} ,
\end{equation}
where $\varepsilon_d$ is the resonator driving strength.
With this, one gets approximately: $\Gamma_{\rm meas} \approx \frac{\tilde{g}_{\parallel}^2}{\kappa/2}$,
so the measurement rate scales quadratically with the dynamical
longitudinal coupling (see also Ref.~\cite{Didier2015PRL}).

By considering the charge qubit example above, \ref{Sec-charge-qubit},
(this is also relevant for the DQD S-T qubit
and for the spin-charge qubit \cite{MiPetta2018N}),
the scaling with the resonator frequency, $\omega_r$ is
\begin{equation}
\frac{\tilde{g}_{\parallel}^{\rm disp}}{\tilde{g}_{\parallel}^{\rm adiabat}}
= \frac{\omega_q}{2 |\omega_q - \omega_r|} .
\label{enhancement_factor_g-parr}
\end{equation}
The enhancement of the measurement rate then scales quadratically with detuning
\begin{equation}
\frac{\Gamma_{\rm meas}^{\rm disp}}{\Gamma_{\rm meas}^{\rm adiabat}}
= \frac{\omega_q^2}{4 (\omega_q - \omega_r)^2 }  \approx \left( 1.6 - 2.5 \right) \, 10^2 ,
\label{enhancement_factor_meas}
\end{equation}
where the estimated numbers are for the same conditions as in Eq.~(\ref{enhancement_factor}).

It is now instructive to compare the measurement rate to the Purcell relaxation rate.
For the Purcell relaxation of
a qubit into the resonator
(for $\kappa \ll \sqrt{(\omega_q - \omega_r)^2 + 4 g_{\perp}^2}$, while $\kappa \gtrsim g_{\perp}$)
one has \cite{SeteGambettaKorotkov2014PRB},
$\Gamma_P \simeq \frac{\kappa g_{\perp}^2}{(\omega_q - \omega_r)^2}$,
i.e. the same scaling, $\sim 1/\Delta^2$ as for the measurement rate.
Therefore, for the purpose of quantum measurements there is no profit of going to the adiabatic regime
as to the Purcell rate supression.

Fortunately, their ratio can be suppressed
\begin{equation}
\frac{\Gamma_P}{\Gamma_{\rm meas}}
= \frac{\kappa^2}{2 g_{\perp}^2} \, \left( \frac{\alpha_c V_{\rm vac}}{\tilde{V}_m} \right)^2
\approx  \frac{\kappa^2}{2 g_{\perp}^2} \, \left(10^{-2} - 10^{-4} \right) ,
\label{Purcell-vs-meas}
\end{equation}
since the external voltage modulation can be made much larger than
amplitude of vacuum voltage fluctuations.

It is worth now to compare to the charge dephasing (due to voltage fluctuations only).
Applying a theoretical model for $1/f$ charge noise
due to gate voltage fluctuations at the dots or tunnel barrier \cite{RuskovTahan2021PRB103}
(see also Ref.~\cite{RussBurkard2015PRB,RussGinzelBurkard2016PRB}),
one gets for a charge qubit at the charge degeneracy point ($\varepsilon=0$)  the rate:
\begin{equation}
\tilde{\Gamma}_{\phi} {\mid_{\varepsilon=0}}
\approx \frac{1}{\hbar}\, \left[ 3 \log \left(\frac{\omega_{\rm UV}}{\omega_{\rm IF}} \right)\,  S_{t_c} \right]^{1/2}
\simeq  \frac{1}{\hbar}\, 0.6\, \mu{\rm eV} ,
\label{one-ef-noise-cdp}
\end{equation}
where the ratio of ultraviolet to infrared cutoff parameters for the $1/f$ spectrum is
$\frac{\omega_{\rm UV}}{\omega_{\rm IF}} \approx 10^6$, and
$S_{t_c} \simeq 10^{-2}\, S_{\varepsilon}$, $S_{\varepsilon} \simeq (1\, \mu{\rm eV})^2$,
are the spectral density constants of the $1/f$ noise associated with the dot gates ($S_{\varepsilon}$)
and tunneling barrier ($S_{t_c}$), extracted from
the
experiment,
see Ref.~\cite{RuskovTahan2021PRB103}.
For the above parameters one estimates dephasing time $1/\tilde{\Gamma}_{\phi} \simeq 1.1\, {\rm ns}$
which is
comparable with the experimentally measured value \cite{MiPetta2018N}.

Out of the charge degeneracy point ($\varepsilon \gg 2 t_c$)
the rate is dominated by the dot gates fluctuations:
\begin{equation}
\tilde{\Gamma}_{\phi} \mid_{\varepsilon \gg 2 t_c}
\approx \frac{1}{\hbar}\,
\left[ \frac{1}{2} \log \left( \frac{\omega_{\rm UV}}{\omega_{\rm IF}} \right)\,  S_{\varepsilon} \right]^{1/2}
\simeq  \frac{1}{\hbar}\, 2.8\, \mu{\rm eV} ,
\label{one-ef-noise-out-of-cdp}
\end{equation}
which corresponds to even ($\approx 5$ times) shorter dephasing time.

This situation considerably improves for a spin-charge qubit, i.e. a charge qubit with
magnetic field gradient between the two dots \cite{HuLiuNori2012PRB,MiPetta2018N}.
For the parameters of the experiment, the measured dephasing rate of the spin-charge qubit
is $\gamma_s \simeq 0.4\, {\rm MHz}$ at $\varepsilon=0$ (c.d.p. at $2 t_c \simeq 11\, {\rm GHz}$), which
corresponds to a 400 times longer dephasing time, $T_s = 0.4\, \mu{\rm s}$.
The physical reason for this is that the qubit levels have opposite spin \cite{MiPetta2018N},
$|0\rangle \equiv |-,\downarrow \rangle$ and $|1\rangle \equiv |-,\uparrow \rangle$,
while an emission of acoustic phonons cannot flip the spin (see, e.g., Ref.~\cite{Yang2013NC}).

For the measured parameters of the experiment of Ref.~\cite{MiPetta2018N}
they have $\kappa \simeq 1.4\, {\rm MHz}$, $\omega_r \simeq 5.85 \, {\rm GHz}$,
spin-charge dipole coupling, $g_s \simeq 1.4 \, {\rm MHz}$,
charge noise dephasing rate, $\gamma_s \simeq 0.4 \, {\rm MHz}$  (c.d.p. at $2 t_c \simeq 11\, {\rm GHz}$),
and relaxation rate $\gamma_1(\varepsilon = 0) = 0.05 \, {\rm MHz}$.
One
can calculate the
measurement rate due to dispersive coupling,
$\delta\omega \simeq \chi_s \equiv \frac{g_s^2}{\Delta}$ and no modulation
\cite{RuskovTahan2019PRB99}:
\begin{equation}
\Gamma_{\rm meas} (\tilde{g}_{\parallel} = 0)
= \frac{(2 \delta\omega)^2 \kappa/2}{\left[ \delta\omega^2 + \kappa^2/4\right]^2} \varepsilon_d^2
\approx 0.1 \, {\rm MHz} \approx 2 \gamma_1 .
\end{equation}
I.e., this measurement rate
is not fast enough
to reach a quantum-limited measurement regime,
as it is comparable both to the charge dephasing rate, $\gamma_s$, and to the relaxation rate, $\gamma_1$.

One can show, however, that with a qubit (gate) modulation with even moderate coupling enhancement
ratio of $\frac{\tilde{g}_{\parallel}}{\delta\omega} \approx 15$ (which is within reach,
see, e.g., Ref.~\cite{CorriganHarptRuskovEriksson2023PRAppl})
one can have a measurement rate:
\begin{equation}
\Gamma_{\rm meas} (\tilde{g}_{\parallel} \neq 0)
\approx \frac{\tilde{g}_{\parallel}^2}{\kappa/2} \approx 50 \, \Gamma_{\rm meas} (\tilde{g}_{\parallel} = 0) ,
\end{equation}
which is considerably stronger.
Thus, with the use of the dynamical longitudinal coupling, $\tilde{g}^{\rm disp}_{\parallel}(\omega_r)$,
in a dispersive regime,
Eq.~(\ref{dispersive-longitudinal-limit}),
one can perform a close to quantum limited continuous measurements of a spin-charge qubit.

\subsection{Implications for geometric quantum gates}

While there are other means of parametric driving \cite{SrinivasaTaylorPetta2023preprint}
aming to obtain entanglement gate between remote spin qubits,
here we perform an estimation based on the geometric phase gates
proposed in Refs.~\cite{OronaYacoby2018PRB,RuskovTahan2021PRB103}.
As shown in Ref.~\cite{RuskovTahan2021PRB103},
in order to obtain high-gate fidelity
one needs to suppress ``dispersive-like'' coupling $\delta\omega$
with respect to the dynamical longitudinal one, $\tilde{g}_{\parallel}$.
E.g., to reach infidelity of $10^{-2} - 10^{-3}$ in the presence of $\delta\omega$
one needs the ratio small, $\frac{\delta\omega}{\tilde{g}_{\parallel}} = 0.011 - 0.035$,
which is within reach, see Eq.~(\ref{ratio_classical-quantum}).

Assuming this ratio is suppressed,
the main source of infidelity in the QD spin system
is that due to the charge noise, see Ref.~\cite{RuskovTahan2021PRB103}.
For a two-qubit controlled $\pi$-phase gate the infidelity reads \cite{RuskovTahan2021PRB103}:
\begin{equation}
\delta\varepsilon^{2Qb}_{\phi,1/f} = \frac{8}{10} \left(\tilde{\Gamma}_{\phi}\, t_{\pi} \right)^2 ,
\label{2Qb_one-over-ef-infidelity}
\end{equation}
where the $\pi$-phase gate time, $t_{\pi} = \frac{\pi \sqrt{2}}{\tilde{g}_{\parallel}}$
(for equal couplings, $\tilde{g}_{\parallel}^{(1)} = \tilde{g}_{\parallel}^{(2)}$),
is inversely proportional to the dynamical longitudinal coupling, $\tilde{g}_{\parallel}$.
Assuming the experimental charge dephasing rate for a spin-charge qubit \cite{MiPetta2018N}
$(\tilde{\Gamma}_{\phi}^{\rm exp} \equiv \gamma_s = 0.41 \, {\rm MHz}$,
such that
%
$\left( \tilde{\Gamma}_{\phi}^{\rm exp} \right)^{-1} \simeq 400\, {\rm ns}$),
to get infidelity of the level of $10^{-1} - 10^{-3}$
one needs $\tilde{g}_{\parallel}/2\pi \approx 5 - 50 \, {\rm MHz}$,
which is reachable due to
an
enhancement at the charge degeneracy point, see Eqs.~(\ref{ratio_curvatures})
and at the dispersive regime (\ref{enhancement_factor}).
Indeed, already in the current experiment \cite{MiPetta2018N}
the dispersive coupling
can reach $\chi_s \simeq 0.14 - 0.55 \, {\rm MHz}$ and the enhancement factor, Eq.~(\ref{ratio_classical-quantum}),
can make the dynamical longitudinal coupling in a dispersive regime
as large as $\tilde{g}_{\parallel} \approx 2 - 66 \, {\rm MHz}$.

\section{Conclusion}

In this paper we have derived effective interaction Hamiltonians, generically called ``dispersive-like''
and dynamical longitudinal
one,
for an $n$-level   
atom
coupled to a superconducting resonator.
These interaction Hamiltonians replace the ``original'' electric dipole interaction,
$-\hat{\overrightarrow{d}} \cdot \delta\overrightarrow{E}_\lambda(t)$,
in a situation when the frequency of the e.m. field $\overrightarrow{E}_\lambda(t)$
is relatively small to create excitations in the $n$-level system.
These Hamiltonians are diagonal in the system eigenlevels.
The ``dispersive-like'' Hamiltonian (time-independent)  is of ``energy-energy'' type,
${\cal H}_{\delta\omega} \sim \delta\omega_i \, |i\rangle\langle i| a^\dagger a$.
The dynamical longitudinal (time-dependent) Hamiltonian  is of ``energy-field'' type,
${\cal H}_{\parallel} \sim  \tilde{g}_{\parallel,i} \, |i\rangle\langle i| (a + a^\dagger) \cos( \omega_m t)$,
and appears due to periodic voltage modulation of a qubit gate.

It is also worth mentioning that (for a two-level system, and making the replacements:
$a^\dagger a \to \sigma_z^{\rm target}$ and $a + a^\dagger \to 2\sigma_x^{\rm target}$)
the ``dispersive-like'' interaction resembles that of a residual $ZZ$ two-qubit term,
$\sim \sigma_z\, \sigma_z^{\rm target}$,
and the dynamical longitudinal one resembles that of a cross-resonance qubit-qubit interaction,
$\sim \sigma_z\, \sigma_x^{\rm target} \cos( \omega_m t)$, driven at the frequency of the ``target'' qubit
(see, e.g., Refs.~\cite{RigettiDevoret2010PRB,TriphatiKhezriKorotkov2019PRA}).

The derivation of these effective Hamiltonians is presented in two ways.
First, in a more heuristic way, we treat the quantized e.m. field of the resonator
as a time-dependent classical field (with frequency $\omega_r$).
Thus, both classical and quantum field  perturbations are treated in
a kind of time-dependent perturbation theory in second order, to derive the effective interactions.
Secondly, in a more formal way, we consider
a time-dependent Schrieffer-Wolff transformation based
on the lab frame (time-independent) photon operators, $\hat{a}, \hat{a}^\dagger$,
to reproduce the same results.

The effective interactions reproduce previous results in limiting cases.
E.g., in the absence of qubit gate modulation
(and dropping off a polarizability contribution),
the ``dispersive-like'' Hamiltonian is
that of Zhu et al. \cite{Zhu:2013}.
With qubit gate modulation, we reproduce the dynamical longitudinal interaction
of a transmon \cite{Touzard:2019}
in the dispersive regime,
for example.

As a side note, we mention that our derivations can be equally applied
to a qubit that is coupled inductively to a SC resonator, e.g.
in a flux qubit \cite{FriedmanLukens2000N,DevoretWallraffMartinis2004preprint},
a charge-flux qubit \cite{Zorin-ZhETF2004},
or Andreev qubit \cite{HaysDevoret2020NP}.
In this case one should consider the
system's
energy curvatures vs. magnetic flux,
which constitute the quantum (Josephson) reverse inductances,
$L_{k,J}^{-1} \propto \frac{\partial^2 E_k}{\partial \Phi^2}$,
compare with, e.g. \cite{Zorin-ZhETF2004,HaysDevoret2020NP},
and consider the quantized flux, $\hat{\Phi}_r$,
of the resonator instead of the quantized voltage $\hat{V}_r$.

We have considered in this paper both the dispersive and adiabatic (ultra-dispersive)
regimes. In the latter, the modulation frequencies are much smaller then any qubit transition
frequencies, $\omega_{kl}$.
For both effective interactions we have shown that in this limit ($\omega_r,\omega_m \ll \omega_{kl}$)
the effective Hamiltonians are expressed through the energy curvatures of the levels,
$\frac{\partial^2 E_i}{\partial V_G^2}$, which is a non-trivial consequence
of a (low energy) quantum-mechanical sum rule \cite{LevyYeyati2020PRL}.
Thus, we exactly reproduce in a new way our  results
for the effective Hamiltonians, ${\cal H}_{\delta\omega}$, ${\cal H}_{\parallel}$, in the adiabatic limit,
that was previously derived via Taylor expansion of the energy levels with respect to
a voltage parameter \cite{RuskovTahan2017preprint1,RuskovTahan2019PRB99}.

As an application of the general theory, we consider several examples of quantum dot qubits
including a charge qubit,
a DQD Singlet-Triplet qubit, and a Transmon.
The charge qubit example is relevant to a recent experiment \cite{CorriganHarptRuskovEriksson2023PRAppl}
demonstrating detailed observation
of the curvature couplings, $\delta\omega$, $\tilde{g}_{\parallel}$
in the adiabatic
regime.
The DQD S-T qubit example will be relevant to a recent experiment \cite{BottcherYacobyNatComm2022}
on parametric longitudinal coupling
or its extension.
We have also performed a crude estimations relevant to the spin-charge qubit
of the J.R. Petta's group \cite{HuLiuNori2012PRB,MiPetta2018N},
showing that using the dynamical longitudinal coupling (via gate modulation)
can significantly increase the quantum measurement rate
(see e.g. Refs.~\cite{RuskovTahan2017preprint1,RuskovTahan2019PRB99}) so that the system can approach
a quantum-limited measurement regime, a prerequisite for interesting quantum measurement experiments,
e.g. performing continuous quantum feedback control \cite{RuskovKorotkov2002PRB66,KorotkovSiddiqi2012N},
entanglement-by-continuous joint measurement \cite{RuskovKorotkov2003PRB67,RisteDiCarlo2013N,KorotkovSiddiqi2014PRL},
and others,
however,
with spin qubits.
\\[3mm]

\noindent
{\bf Acknowledgments.}
We thank Mark Eriksson and Mark Friesen for useful
conversations and stimulating discussions concerning
a part of this work.

\appendix

\section{Derivation of the effective Hamiltonian, Eq.~(\ref{H-eff}),
via time-dependent Schrieffer-Wolff transformation}
\label{app-A: Schrieffer-Wolff-PT}

%
An $n$-level atom and the e.m. field mode of frequency $\omega_r$ are represented by
the Hamiltonian ${\cal H}_0$, and interact via the (time-dependent)
interaction $\hat{V}(t)$,
${\cal H}(t) = {\cal H}_0 + \hat{V}(t)$
(below $\hbar=1$).
In the case of a dipole interaction with external gate voltage modulation
one gets $\hat{V}(t) \equiv {\cal H}_{\rm dipole}$, Eq.~(\ref{V-dipole-transverse-coupl}).
In what follows, we consider only the non-diagonal contributions, $\hat V_{\rm nd}(t)$,
as the diagonal contributions
are suppressed in the rotating wave approximation (RWA):
\begin{eqnarray}
&& {\cal H}_0 = \sum_k \omega_k | k \rangle \langle k | +  \omega_r \hat{a}^{\dagger} \hat{a}
\label{H_0}
\\
&& \hat{V}_{\rm nd}(t)
= \sum_{k,l,k\neq l} g_{kl} |k \rangle \langle l |\, \left( \hat{a}^{\dagger} + b^*(t) + \hat{a} + b(t)\right)
\label{transition-dipole-interaction}
\end{eqnarray}
(note that $g_{kl} = g_{lk}^{*}$).
One can obtain an effective Hamiltonian ${\cal H}_{\rm eff}(t)$,  diagonal in the atom index,
by applying a time-dependent unitary transformation $U_1(t) = \exp[-S_1(t)]$
(time-dependent Schrieffer-Wolff transformation \cite{MagesanGambetta2020PRA,GoviaKamal2022PRA}).
It can be shown that to eliminate the off-diagonal terms in the atom index
in the next order in perturbation theory (PT) the operator $S_1(t)$
needs to satisfy the equation \cite{GoviaKamal2022PRA} :
\begin{equation}
i\, \frac{\partial S_1(t)}{\partial t} + [S_1(t),{\cal H}_0] + V_{\rm nd}(t) = 0
\label{time-dependent-SF-equation} .
\end{equation}
The operator $S_1^{\dagger}(t) = - S_1(t)$ is anti-Hermitian
and will be searched in the form
\begin{equation}
S_1(t) = \sum_{l,l'} \, S_{l'l}(t) |l' \rangle \langle l | \left( \hat{a}^{\dagger} + b^*(t) \right) - {\rm H.c.}
\label{S-F-operator-S} ,
\end{equation}
where $S_1(t)$ is off-diagonal, i.e. $S_{ll}=0$.

The commutator $[S_1(t),{\cal H}_0]$ is obtained using the standard
relations,
$\left[|l'\rangle \langle l|, |m\rangle \langle m'|\right]
= |l'\rangle \langle m'|\,\delta_{lm} - |m \rangle \langle l|\, \delta_{m'l'}$,
and $\left[\hat{a},\hat{a}^{\dagger}\right] = 1$.
One gets:
\begin{eqnarray}
&& [S_1(t),{\cal H}_0]
= \sum_{l,l} \, \left\{ S_{l'l}(t) |l' \rangle \langle l | \,
\left[ \omega_{ll'} \left( \hat{a}^{\dagger} + b^*(t) \right)  -  \omega_r \hat{a}^{\dagger} \right] \right.
\nonumber\\
&& \qquad\qquad
\left. { } + S_{l'l}^*(t) |l \rangle \langle l' | \,
\left[ \omega_{ll'} \left( \hat{a} + b(t) \right)  -  \omega_r \hat{a} \right] \right\} ,
\label{commutator_S1_H0}
\end{eqnarray}
where $\omega_{ll'} \equiv \omega_l - \omega_{l'}$.
One now substitute Eq.~(\ref{S-F-operator-S}) for $S_1(t)$ into Eq.~(\ref{time-dependent-SF-equation})
and obtain equations at the different operator structures.
E.g., at the structures $| l' \rangle \langle l |\, \hat{a}^{\dagger}$ and
$| l' \rangle \langle l |\, \hat{a}$ one gets, respectively:
\begin{eqnarray}
&& i\, \frac{\partial S_{l'l}(t)}{\partial t} + S_{l'l}\, \left(\omega_{ll'} - \omega_r \right) + g_{l'l} = 0
\label{S_dot_equation}
\\
&& -i\, \frac{\partial S_{ll'}^*(t)}{\partial t} + S_{ll'}^*\, \left(\omega_{l'l} - \omega_r \right) + g_{l'l} = 0
\label{S_dot_cc_equation}
\end{eqnarray}
%
One should note that the equation obtained at the structure $| l' \rangle \langle l |$ is not independent, but
is a linear combination
of Eq.~(\ref{S_dot_equation}) and Eq.~(\ref{S_dot_cc_equation}).
For time-independent dipole couplings, $g_{l'l}$, one gets the solutions:
\begin{eqnarray}
&&  S_{l'l} = - \frac{g_{l'l}}{ \omega_{ll'} - \omega_r }
\label{S_solution}
\\
&& S_{ll'}^* = - \frac{g_{l'l}}{ \omega_{l'l} - \omega_r } .
\label{S_cc_solution}
\end{eqnarray}
Substituting these results into Eq.~(\ref{S-F-operator-S})
one obtains $S_1(t)$, Eq.~(\ref{S_1_solution}) of the main text.

Using the result for $S_1(t)$, the effective Hamiltonian in the
dispersive/adiabatic regime is calculated
from the commutator:
\begin{equation}
{\cal H}_{\rm eff}\mid_{\rm RWA} =  \frac{1}{2} \, [S_1(t), \hat{V}_{\rm nd}(t)]
\label{H_eff_commutator}
\end{equation}
in the RWA.
Calculating the commutator we have neglected
terms of the form,
$\left( \hat{a}^{\dagger} + b^*(t) \right)^2$, $\left( \hat{a} + b(t) \right)^2$, in a RWA.

Thus, we recover the effective Hamiltonian, Eq.~(\ref{H-eff}) of the main text,
that was obtained via
a time-dependent PT.

\section{Quantum-mechanical sum rule for the polarizability matrix element}
\label{app-B: Low-energy-sum-rule}

For completeness, here we present  a detailed derivation of the quantum-mechanical sum rule,
Eq.(\ref{QuMech-sum-rule}) of the main text.
essentially following Ref.~\cite{LevyYeyati2020PRL}.

To express the diagonal matrix element of the polarizability in the energy eigenbasis,
$\langle \psi_i | \frac{\partial^2 {\cal H}_{\rm qb}}{\partial V^2} | \psi_i\rangle$,
one differentiates the Hellmann-Feynman relation,
$\frac{\partial E_i}{\partial V} = \langle \psi_i | \frac{\partial {\cal H}_{\rm qb}}{\partial V} | \psi_i\rangle$,
with respect to a suitable voltage parameter $V$
to obtain:
\begin{equation}
\frac{\partial^2 E_i}{\partial V^2} =
\langle \frac{\partial \psi_i }{\partial V}| \frac{\partial {\cal H}_{\rm qb}}{\partial V} | \psi_i\rangle
+ \langle \psi_i | \frac{\partial^2 {\cal H}_{\rm qb}}{\partial V^2} | \psi_i\rangle
+ \langle \psi_i | \frac{\partial {\cal H}_{\rm qb}}{\partial V} | \frac{\partial \psi_i}{\partial V}\rangle .
\label{2nd-derivative-Feynman-Hellmann}
\end{equation}
It is convenient to introduce the i-th level Green's function, $G_i \equiv \frac{1}{E_i - {\cal H}_{\rm qb}}$.
Differentiating the relation: $G_i^{-1} | \psi_i\rangle \equiv (E_i - {\cal H}_{\rm qb}) | \psi_i\rangle = 0$
one gets:
\begin{equation}
| \frac{\partial \psi_i}{\partial V}\rangle = - G_i \frac{ \partial(G_i^{-1})}{\partial V} | \psi_i\rangle .
\label{psi-prime}
\end{equation}

To evaluate the 3rd term in Eq.~(\ref{2nd-derivative-Feynman-Hellmann}) one substitutes in it
Eq.~(\ref{psi-prime}) to obtain:
\begin{eqnarray}
&& \langle \psi_i | \frac{\partial {\cal H}_{\rm qb}}{\partial V} | \frac{\partial \psi_i}{\partial V}\rangle =
\nonumber\\
&& \qquad { } = - \sum_{j,k}\, \langle \psi_i | \frac{\partial {\cal H}_{\rm qb}}{\partial V} | \psi_j\rangle \,
\langle \psi_j | G_i | \psi_k\rangle \, \langle \psi_k | \frac{ \partial(G_i^{-1})}{\partial V} | \psi_i\rangle ,
\qquad
\label{3rd-term}
\end{eqnarray}
where we have inserted the completeness condition, $\sum_j \, | \psi_j \rangle \langle \psi_j | = I$.
For the last multiplier of Eq.~(\ref{3rd-term}) one gets:
\begin{equation}
\langle \psi_k | \frac{ \partial(G_i^{-1})}{\partial V} | \psi_i\rangle =
\delta_{ki} \, \frac{\partial E_i}{\partial V} -
\langle \psi_k | \frac{\partial {\cal H}_{\rm qb}}{\partial V} | \psi_i\rangle .
\label{last-multiplier}
\end{equation}
By differentiating the identities,
$\langle \psi_k | {\cal H}_{\rm qb} | \psi_i\rangle = \delta_{ki} E_i$
and $\langle \psi_k | \psi_i\rangle = \delta_{ki}$,
one gets the simple relations:
\begin{eqnarray}
&& \langle \psi_k | \frac{\partial {\cal H}_{\rm qb}}{\partial V} | \psi_i\rangle
= \delta_{ki} \, \frac{\partial E_i}{\partial V}
- \langle \frac{\partial \psi_k}{\partial V} | \psi_i \rangle \, E_i
- \langle \psi_k | \frac{\partial \psi_i}{\partial V} \rangle \, E_k  \qquad
\\
&& \langle \frac{\partial \psi_k}{\partial V} | \psi_i \rangle
 + \langle \psi_k | \frac{\partial \psi_i}{\partial V} \rangle = 0 \,  ,
\label{simple_relations}
\end{eqnarray}
and combining them one gets:
\begin{equation}
\langle \psi_k | \frac{\partial {\cal H}_{\rm qb}}{\partial V} | \psi_i\rangle
= \delta_{ki} \, \frac{\partial E_i}{\partial V}
+ (E_i - E_k)\, \langle \psi_k | \frac{\partial \psi_i}{\partial V}\rangle .
\label{non-diagonal-Feynman-Hellmann}
\end{equation}
Substituting Eqs.~(\ref{non-diagonal-Feynman-Hellmann}) and (\ref{last-multiplier})
into Eq.~(\ref{3rd-term}) and taking into account that
$\langle \psi_j | G_i | \psi_k\rangle = \frac{1}{E_i - E_k}\, \delta_{jk}$
one finally obtains for 3rd term of
Eq.~(\ref{2nd-derivative-Feynman-Hellmann}):
\begin{equation}
{\rm (B3)} \equiv
\langle \psi_i | \frac{\partial {\cal H}_{\rm qb}}{\partial V} | \frac{\partial \psi_i}{\partial V}\rangle
= - \sum_{j\neq i}\,
\frac{|\langle \psi_i|\frac{\partial {\cal H}_{\rm qb}}{\partial V} |\psi_j \rangle|^2}{E_j - E_i}
\end{equation}

The first term in Eq.~(\ref{2nd-derivative-Feynman-Hellmann}) is expressed via the third one,
(B3),
since $\frac{\partial {\cal H}_{\rm qb}}{\partial V}$ is Hermitean:
\begin{equation}
\langle \frac{\partial \psi_i }{\partial V}| \frac{\partial {\cal H}_{\rm qb}}{\partial V} | \psi_i\rangle
=
\left(\langle \psi_i | \frac{\partial {\cal H}_{\rm qb}}{\partial V} | \frac{\partial \psi_i}{\partial V}\rangle\right)^{*} .
\label{1st-term}
\end{equation}
Substituting this
in Eq.~(\ref{2nd-derivative-Feynman-Hellmann})
one recovers the quantum-mechanical sum rule,
Eq.(\ref{QuMech-sum-rule}), for the
polarizability matrix element:
\begin{equation}
\langle \psi_i | \frac{\partial^2 {\cal H}_{\rm qb}}{\partial V_G^2} | \psi_i\rangle
= \frac{\partial^2 E_i}{\partial V_G^2}
+ 2 \sum_{j\neq i}
\frac{|\langle \psi_i | \frac{\partial {\cal H}_{\rm qb}}{\partial V_G} |\psi_j\rangle |^2}{E_j - E_i} .
\label{QuMech-sum-rule-app}
\end{equation}

The mere purpose of this sum rule is to establish the proper limit
for ${\cal H}_{\delta\omega}$ and ${\cal H}_{\parallel}$, Eqs.~(\ref{dispersive_Hamiltonian_general})
and (\ref{dynamical_longitudinal_Hamiltonian_general})
in the adiabatic regime, when they are expressed via the energy curvature w.r.t. voltage
(i.e., quantum capacitance) effective Hamiltonians,
Eqs.~(\ref{qudit-dispersive-like}), (\ref{qudit-dynamical-longitudinal}),
and Eqs.~(\ref{adiabatic-limit-delta-omega}), (\ref{adiabatic-limit-g}).

\section{Dipole matrix elements of the TQD exchange only qubit at or around the
full sweet spot}
\label{app-C: Dipole-m-e-TQD-qubit}

Here we briefly sketch the calculation of the transition dipole moments
of the TQD system,
where the higher excited (doubly-occupied) states contributions
will dominate in the effective adiabatic interactions at or around the
full sweet spot (SOP).
Some details of the calculations can be found in Ref.~\cite{RuskovTahan2019PRB99}.
We repeat some of the results and definitions from that reference
for the sake of completeness.

The TQD Hamiltonian and the dipole interaction are
formulated  \cite{RuskovTahan2019PRB99}
in the charge basis of the 6 states of spin projection $S_z = +1/2$, namely
$| 1\rangle
    = \frac{1}{\sqrt{2}}
\left( | \uparrow_1\uparrow_2\downarrow_3\rangle - | \downarrow_1\uparrow_2\uparrow_3 \rangle \right)$,
$| 2 \rangle =
-\frac{1}{\sqrt{6}}
\left( | \uparrow_1\uparrow_2\downarrow_3 \rangle + | \downarrow_1\uparrow_2\uparrow_3 \rangle -
2 | \uparrow_1\downarrow_2\uparrow_3 \rangle\right)$
[qubit subspace with charge configuration $(1,1,1)$],
and
the 4 highly gapped charge states ($E_{\rm gap} \approx U_{\rm charge}$), i.e.
$| 3 \rangle = (201) = |S(2_1,0_2)\rangle |\uparrow_3 \rangle$,
$| 4 \rangle = (102) = |\uparrow_1 \rangle | S(0_2,2_3) \rangle$,
$| 5 \rangle = (120) = |\uparrow_1 \rangle | S(2_2,0_3) \rangle$,
$| 6 \rangle = (021) = | S(0_1,2_2) \rangle |\uparrow_3 \rangle$.
The TQD system Hamiltonian has diagonal energies,
$E_1 = E_2 = 0$,
$E_3 = \varepsilon_v - \varepsilon_m + \tilde{U}_1$,
$E_4 = - \varepsilon_v - \varepsilon_m + \tilde{U}_3$,
$E_5 = \varepsilon_v + \varepsilon_m + \tilde{U}_2$,
$E_6 = - \varepsilon_v + \varepsilon_m + \tilde{U}_2^{'}$,
and
off-diagonal tunneling matrix elements
that
are linear in the left or right tunneling
amplitudes, $t_l$, $t_r$, that couple only the qubit subspace,
$\{|1\rangle,|2\rangle\}$ to the upper gapped states \cite{RuskovTahan2019PRB99}.
The dipole interaction in the charge basis is diagonal
and assuming
the coupling to the SC resonator is through the middle dot 2 via the $V_2$ gate voltage
(we neglect small corrections due to capacitance coupling between the dots),
one gets:
\begin{eqnarray}
&& {\cal H}_{\rm dipole} = 2 \hbar g_0 (\hat{a} + \hat{a}^\dagger) \,\  \hat{D} ,
\label{dipole-Hamiltonian}\\
&&  \qquad  \hat{D} \equiv  {\rm diag} [1,\, 1,\, 0,\, 0,\, 2,\, 2] .
\nonumber  \ \ \
\end{eqnarray}
This
is obtained using the linear response approach \cite{RuskovTahan2019PRB99}.
In the above energies of the excited states we have used the notations of
the two gate voltage (energy) detunings,
$\varepsilon_v \equiv e(V_3 - V_1)/2$,
and $\varepsilon_m \equiv e[(V_3 + V_1)/2 - V_2]\equiv e V_m$.
Also, the charging energies, $\tilde{U}_i \approx U_{\rm charge}$, are defined as the energy costs
to go from the $(1,1,1)$ configuration to a configuration where the $i$-th dot is doubly occupied,
e.g., $\tilde{U}_1$ is the energy cost
for transition from
$(1,1,1)$ to $(2,0,1)$, etc.

One first performs a
(static)
Schrieffer-Wolff transformation \cite{BirPikusBook,SchriefferWolff}
that brings the TQD Hamiltonian to a block-diagonal form, decoupling the qubit subspace from the
highly gapped 4 states.
The qubit block (Hamiltonian) takes the well known form
in the transformed basis,
\begin{eqnarray}
&& {\cal H}^{\rm TQD}_{\rm q} = - J(\epsilon_v,\epsilon_m)
+ \frac{J(\epsilon_v,\epsilon_m)}{2} \tilde{\sigma}_z
- \frac{\sqrt{3}}{2} j(\epsilon_v,\epsilon_m)\, \tilde{\sigma}_x
\qquad
\label{TQD-Hamiltonian-BirPikus}\\
&& \qquad\quad { } = - J(\varepsilon_v,\varepsilon_m)  - \frac{E_q(\epsilon_v,\epsilon_m)}{2} \sigma_z,
\end{eqnarray}
with the exchange energies $J\equiv (J_l + J_r)/2$, $j\equiv (J_l - J_r)/2$
and a qubit splitting $E_q = \sqrt{J_l^2 + J_r^2 -J_l J_r}$,
where the
left and right exchange energies are given as,
\begin{eqnarray}
&& J_l(\varepsilon_v,\varepsilon_m) = 2 t_l^2 \left[ \frac{1}{\varepsilon_v - \varepsilon_m + \tilde{U}_1}
+ \frac{1}{- \varepsilon_v + \varepsilon_m + \tilde{U}_2^{'}} \right]
\label{J_left}\\
&& J_r(\varepsilon_v,\varepsilon_m) = 2 t_r^2 \left[ \frac{1}{\varepsilon_v + \varepsilon_m + \tilde{U}_2}
+ \frac{1}{- \varepsilon_v - \varepsilon_m + \tilde{U}_3} \right] .
\label{J_right}
\end{eqnarray}
In Eq.~(\ref{TQD-Hamiltonian-BirPikus}) the
diagonalization is
further
performed by a unitary transformation
\begin{equation}
U_{\rm qb} =
\left(
\begin{array}{cc}
\cos (\eta/2)\ , & \sin (\eta/2)\\
- \sin (\eta/2)\ , & \cos (\eta/2)
\end{array}
\right)
\end{equation}
where
\begin{equation}
\eta/2 = \arccos\left[\frac{1}{\sqrt{2}}\, \left(1 - \frac{J}{\sqrt{J^2 + 3 j^2}}  \right)^{1/2} \right] .
\end{equation}
Thus,
the qubit eigenstates $|+\rangle$, $|-\rangle$
obtain energy curvature with respect to
voltage
detunings, $\epsilon_v$, $\epsilon_m$
due to the higher levels.
The other block of the 4 highly gapped states (at or around the full sweet spot)
remains approximately diagonal, since the corrections to the diagonal energies,
$E_l \approx U_{\rm charge},\ l=3,4,5,6$,
are of the order of $\sim t_{l,r}^2/U_{\rm charge}$,
i.e highly suppressed for $t_{l,r} \sim 5 - 10\, {\rm GHz}$ and $U_{\rm chrage} \sim 200 -300 \, {\rm GHz}$.

By performing the same transformations as above to the dipole Hamiltonian,
one obtains the qubit dipole coupling, $g_{-,+}$, as \cite{RuskovTahan2019PRB99}:
\begin{equation}
g_{-,+} \equiv g_{\perp} = - g_0 \frac{\sqrt{3}}{4 E_q}
\left[ \frac{\partial J_r}{\partial \epsilon_m} J_l - \frac{\partial J_l}{\partial \epsilon_m} J_r \right] .
\label{g-transverse-TQD}
\end{equation}
Note that at the full sweet spot (SOP) when the detunings take the values\cite{AEON2016},
\begin{equation}
\varepsilon_v^0 = \frac{1}{4}( - \tilde{U}_1 + \tilde{U}_2^{'} - \tilde{U}_2 + \tilde{U}_3),\
\varepsilon_m^0 = \frac{1}{4}(\tilde{U}_1 - \tilde{U}_2^{'} - \tilde{U}_2 + \tilde{U}_3) ,
\label{sweet-spot}
\end{equation}
the qubit dipole coupling is zero, $g_{-,+}=0$ \cite{RuskovTahan2019PRB99}.
It is also zero in the symmetric situation, $t_l = t_r$, $\tilde{U}_1 = \tilde{U}_3$,
$\tilde{U}_2^{'} = \tilde{U}_2$, and $\varepsilon_v = \varepsilon_v^0 = 0$ is at sweet spot value
while $\varepsilon_m$ is arbitrary \cite{RuskovTahan2019PRB99}.

For the dipole couplings of the qubit levels to the upper highly gapped levels,
one obtains (up to a factor of $2 g_0$):
%
\begin{eqnarray}
&& g_{-,3} = \frac{\sqrt{2}\, t_l \, \sin(\alpha_{+})}{\varepsilon_v - \varepsilon_m + \tilde{U}_1}, \ \
 g_{-,4} = \frac{\sqrt{2}\, t_r \, \sin(\alpha_{-})}{-\varepsilon_v - \varepsilon_m + \tilde{U}_3} \quad
\\
&& g_{-,5} = -\frac{\sqrt{2} t_r \sin(\alpha_{-})}{\varepsilon_v + \varepsilon_m + \tilde{U}_2}, \ \
g_{-,6} = -\frac{\sqrt{2} t_l \sin(\alpha_{+})}{-\varepsilon_v + \varepsilon_m + \tilde{U}_2^{'}} \qquad\quad
\label{g_-l}
\end{eqnarray}
and
\begin{eqnarray}
&& g_{+,3} = \frac{\sqrt{2} t_l \cos(\alpha_{+})}{\varepsilon_v - \varepsilon_m + \tilde{U}_1}, \ \
g_{+,4} = \frac{\sqrt{2} t_r \cos(\alpha_{-})}{-\varepsilon_v - \varepsilon_m + \tilde{U}_3} \quad
\\
&& g_{+,5} = -\frac{\sqrt{2} t_r \cos(\alpha_{-})}{\varepsilon_v + \varepsilon_m + \tilde{U}_2}, \ \
g_{+,6} = -\frac{\sqrt{2} t_l \cos(\alpha_{+})}{-\varepsilon_v + \varepsilon_m + \tilde{U}_2^{'}} \qquad\quad
\label{g_+l}
\end{eqnarray}
where
$\alpha_{\pm} \equiv \frac{\pi}{6} \pm \frac{\eta}{2}$.

These dipole
couplings
are used in the main text, Sec.~\ref{Sec-TQD exchange only qubit}.
They are relevant in a range of detunings,
$\varepsilon_v \approx \varepsilon_v^0$, $\varepsilon_m \approx \varepsilon_m^0$
at or around the full sweet spot.

Calculations for the
resonant exchange (RX)
regime \cite{Srinivasa2016PRB} are analogous to the above,
but not performed here.
In the RX regime (approaching a c.d.p.) some of the energy denominators become small
and their contribution are enhanced. Then the block of 4 upper states need to be diagonalized
as well, in order to obtain the dipole elements to the upper states, $g_{-,l}$, $g_{+,l}$.


%

\end{document}